\documentclass[aps,showpacs,superscriptaddress]{revtex4}
\usepackage{graphicx}
\begin{document}
\title{Microscopic analysis of  shape-phase transitions
in  even-even $N\sim 90$ rotating nuclei}
\author{J. Kvasil}
\affiliation{Institute of Particle and Nuclear Physics, Charles
University, V.Hole\v sovi\v ck\'ach 2, CZ-18000 Praha 8, Czech Republic}
\author{R. G.~Nazmitdinov}
\affiliation{Departament de F{\'\i}sica,
Universitat de les Illes Balears, E-07122 Palma de Mallorca, Spain}
\affiliation{Bogoliubov Laboratory of Theoretical Physics,
Joint Institute for Nuclear Research, 141980 Dubna, Russia}
\date{\today}
\begin{abstract}
We study in cranked Nilsson plus random phase approximation
shape transitions in fast rotating nuclei undergoing backbending, more 
specifically $^{156}$Dy and $^{162}$Yb. We found that a backbending in 
$^{156}$Dy is correlated with the disappearance of the
collective, positive signature $\gamma$-vibrational mode in the rotating frame,
and, a shape transition (axial$\rightarrow$ nonaxial) is accompanied with 
a large acquiring  of the $\gamma$ deformation. We show that such a shape 
transition  can be considered as a phase transition of the first order.
In $^{162}$Yb the quasiparticle alignment dominates in the backbending and 
the shape transition (axial$\rightarrow$ nonaxial) is accompanied with a 
smooth transition from zero to nonzero values of the $\gamma$ deformation. 
We extend the classical Landau theory for rotating nuclei and  show 
that the backbending in $^{162}Yb$ is identified with the second order phase 
transition. A description of spectral and decay properties of the yrast states 
and low-lying excitations demonstrates a good agreement between 
our results and experimental data.
\end{abstract}
\maketitle

\section{Introduction}

Backbending is a paradigm of structural changes in a nucleus under rotation.
A sudden increase of a nuclear moment of inertia in yrast rotational band
at some critical angular momentum or rotational frequency, discovered few decades 
ago \cite{Jon}, continues to attract a considerable attention.  There is a general 
persuasion  that this phenomenon is a result of the rotational alignment of angular 
momenta of a nucleon pair occupying a high-j intruder orbital near the Fermi surface  
(see textbooks \cite{RS,Rag} and references therein). However,
it is understood just recently, that this point of view
may obscure different mechanisms if applied to nuclei with relatively small 
axial deformation at zero spin. In particular, Regan {\it et al} \cite{Re} proposed 
that the backbending observed in a number Cd, Pd and Ru nuclei can be interpreted 
as a change from vibrational to rotational structure. In our preliminary 
report \cite{JK1}, we proposed to consider the backbending in $N\sim90$ as a result of the 
transition from axially symmetric to nonaxial shape due to a disappearance of 
a collective $\gamma$-vibrational mode in the rotating frame.

The analogy of the backbending phenomenon with a behaviour of superconductors in
magnetic field, noticed in Ref.\onlinecite{mot}, is rising a desire to apply
Landau  theory of phase transitions \cite{Lan} to nuclei.
We recall that Landau's theory deals with second order phase transitions, when different
macroscopic phases become indistinguishable at the transition point.
Whether it takes place for rotating nuclei is an open question.
Nuclei are finite systems and phase transitions should be
washed out by quantum fluctuations. 
Nevertheless, long ago Thouless \cite{t22} 
proposed to distinguish two kinds of "phase transitions" even for nuclei.
Such phase transitions may be connected
with shape transitions, for example, from spherical to deformed or
axially deformed to nonaxially deformed shapes.
This idea was put ahead in the analysis of shape transitions in hot
rotating nuclei \cite{Al}.  In this case the statistical treatment of the 
finite-temperature mean field description had provided a justification for
an application of the Landau theory for nuclei. 
Within this approach, a simple rules for different shape-phase transitions
were found as a function of angular momentum and temperature.

Recently, quantum phase transitions, that occur at {\it zero temperature} 
as a function of some nonthermal control parameter, attract a considerable 
attention in various branches of many-body physics, starting from low-dimensional 
systems \cite{Sad} to atomic nuclei and molecules \cite{Zam}. 
Until the present a major activity in the study of shape-phase transitions  
for nuclei in the ground state at zero temperature is carried out 
within the interacting boson model (IBM) (see, for example, 
Refs.\onlinecite{Jo02,cej} and references  therein). 
The model naturally incorporates different symmetry limits associated with specific 
nuclear properties  \cite{nat}. While the IBM can be easily extended to a thermodynamical limit
$N \rightarrow \infty $, which is well suitable for the study of phase transitions,
the analysis is rather oversimplified. For example, the model does not take into 
account the interplay between single-particle and collective 
degrees of freedom in even-even nuclei. 
A general trend found for the ground shape transitions is less affected 
by this interplay. However, it may be crucial for the study of 
quantum phase transitions in rotating nuclei, where statical and dynamical 
properties are coupled.
As we will see below, the above interplay determines the type of quantum 
shape-phase transitions in rotating nuclei. It elucidates also
the behaviour of low-lying excitations, specifically related to the  shape transitions at
high spins. Among such modes are $\gamma$-excitations and wobbling excitations, 
which are related to the nonaxial shapes.

The nuclear shell model (SM) treats the single-particle (s.p.) and collective degrees
of freedom equally and appears to be extremely successful in the calculation
of the backbending curve in light nuclei \cite{sm}. However, the drastic increase
of the configuration space for  medium and heavy systems makes the shell model
calculations impossible. In addition, one needs some model consideration to interpret
the SM results. On the other hand, various cranking
Hartree-Fock-Bogoliubov (HFB) calculations (cf \cite{tan,eg,fr}) provide a reliable
analysis of the backbending for medium and heavy systems. 
As a rule, low-lying rotational bands are described within the cranking
model with a principal axis (PAC) rotation. For the PAC rotation each
single-particle (quasiparticle) configuration corresponds to a band of a given
parity and a signature \cite{fra}. In the HFB calculations the backbending is
explained as a crossing of two quasiparticle configurations with different 
mean field characteristics. 

It is well known, however, that a mean field description 
of finite Fermi systems could break spontaneously one of the symmetries of the
exact Hamiltonian, the so-called spontaneously symmetry breaking (SSB)
phenomenon (see Refs.\onlinecite{fra,sat} for a recent review on the SSB effects in
rotating nuclei). Obviously, for finite systems
quantum fluctuations, beyond the mean field approach, are quite important.
The random phase approximation (RPA) being  an efficient tool to
study these quantum fluctuations (vibrational and rotational
excitations) provides also a consistent way to treat broken symmetries.
Moreover, it separates collective excitations associated with each broken
symmetry as a spurious RPA mode and fixes the corresponding inertial parameter.
This was recently demonstrated in fully self-consistent, unrestricted
Hartree-Fock (HF) calculations for other mesoscopic system, two-dimensional
quantum dots with small number of electrons and a parabolic confinement \cite{llor1,llor2}.
For large enough values of the Coulomb interaction-confinement ratio $R_W$ the
HF mean field breaks circular symmetry; the electrons being localized in
specific geometric distributions.  Applying next the RPA it was shown that the broken symmetry
corresponds to appearance of spurious (with a zero energy) RPA mode (Nambu-Goldstone mode).
And this mode  can be associated with a  rotational collective motion of this specific (deformed)
electron configuration (see \cite{llor1,llor2} and references therein),
which is separated from the vibrational excitations. Thus, a self-consistent mean field
calculations combined with the RPA analysis could be useful to reveal structural changes in
a mesoscopic system, i.e., to detect a quantum shape-phase transition.

In contrast to quantum dots, in nuclei, a nucleon-nucleon interaction is less known.
Mean field calculations with effective density dependent nuclear interactions 
such as  Gogny or Skyrme forces or a relativistic mean field approach still do 
not provide sufficiently accurate single-particle spectra 
to obtain a reliable description of experimental characteristics of
low-lying states (cf \cite{afa}). The RPA analysis based on such mean field solutions 
is focused only on the description of various giant resonances in nonrotating nuclei, 
when the accuracy of single-particle spectra near the Fermi level is not 
important (cf \cite{ben}). Furthermore, a practical application of the RPA 
for the nonseparable effective forces in rotating nuclei requires too
large configuration space and is not available yet.
A self-consistent mean field obtained with the aid of phenomenological 
cranked Nilsson or Saxon-Woods potentials and pairing forces is quite competitive 
up to now, from the above point of view. These potentials allow to construct also
a self-consistent residual interaction neglected at the mean field level.
The RPA with a separable multipole-multipole interaction
based on these phenomenological potentials is an effective tool to study
low-lying collective excitations at high spins (cf \cite{KN,nak}).

It was demonstrated recently, in an exactly solvable cranking harmonic oscillator 
model with a self-consistent separable residual interaction \cite{HN02,n2},
that a direct correspondence between the SSB effects of the rotating
mean field and zero RPA modes can be established in a rotating frame if and only 
if mean field minima are found self-consistently. Thus, it is self-evident that 
the analysis of the SSB effects and RPA excitations for realistic potentials 
requires a maximal accuracy of the fulfillment of self-consistency conditions.
In Ref.\onlinecite{JK1},  we  proposed a
practical method to solve almost self-consistently the mean field problem
for the cranked Nilsson model with the pairing forces in order to study
quadrupole excitations  in the RPA.
In the present paper we discuss all the details of our method 
and analyse the backbending in $^{156}$Dy and $^{162}$Yb.
We thoroughly investigate the positive
signature quadrupole excited bands as a function of the angular rotational frequency.
In contrast to the HFB calculations, low-lying excited states in
our approach are the RPA excitations (phonons) built on the vacuum
states. Our vacuum states are yrast line states, i.e., the lowest energy states 
at a given rotational frequency. Note, that RPA phonons describe collective and 
noncollective excitations equally \cite{33,RS}. The rotational bands are composed 
of the states with a common structure (characterized by the same parity, signature 
and connected by strong B(E2) transitions). We will demonstrate that the positive 
signature excitations are closely related to the shape transitions,
that take place in the considered nuclei undergoing backbending.
Hereafter, for the sake of discussion, we call our approach
the CRPA. The validity of our approach will be confirmed by  a remarkable
agreement  between  available experimental data  and our results
for various quantities like kinematical and dynamical moment of inertia,
quadrupole transitions etc.

The paper is organized as follows:  in Section II we review the Hartree-Bogoluibov
approximation for rotating nuclei and discuss mean field results.
 Section III is devoted to the discussion of positive signature RPA excitations
 and their relation to the backbending phenomenon.
The conclusions are finally drawn in Sec.\ IV.

\section{The Mean Field Solution}
\subsection{The Hartree-Bogoliubov approximation}
We start with the Hamiltonian
\begin{eqnarray}
\label{3}
H_{\Omega} \,&=&\, H - \hbar \Omega \hat I_1 \,=\,
H_0 - \sum_{\tau=n,p} \lambda_{\tau} N_{\tau} - \hbar \Omega I_1 + V \nonumber\\
&=&{\tilde H}_0- \hbar \Omega \hat I_1+V 
\end{eqnarray}
The unperturbed term consists of two pieces
\begin{equation}
H_0=\sum_{i} (h_{Nil}(i) + h_{add}(i)).
\end{equation}
The first is the Nilsson Hamiltonian \cite{Rag}
\begin{equation}
\label{4}
h_{Nil} =  \frac{p^2}{2m} + V_{HO} -
2 \kappa \hbar \omega_{00} {\bf l} \cdot {\bf s} -
\kappa \mu \hbar \omega_{00} ({\bf l}^{2} - \langle{\bf l}^{2}\rangle_N),
\end{equation}
where
\begin{equation}
V_{HO} = \frac{1}{2} m (\omega_1^2 x_{1}^2 + \omega_2^2 x_{2}^2 +
\omega_3^2 x_{3}^2)
\label{TriHO}
\end{equation}
is a triaxial harmonic oscillator (HO) potential, whose frequencies satisfy the volume
conserving condition $\omega_{1} \omega_{2} \omega_{3} = \omega_0^3$
($\hbar \omega_0=41 A^{-1/3}$ MeV).
 In the cranking model with the standard Nilsson
potential \cite{Rag} the value of the moment of inertia is largely overestimated 
due to the presence of the velocity dependent $\vec l\,^2$ term. 
This term favours s.p. orbitals with large orbital momenta $l$ and drives
a nucleus to a rigid body rotation too fast. This shortcoming can
be overcome by introducing the additional term.
The second piece of $H_0$ restores  the local
Galilean invariance broken in the rotating coordinate
system and has the form \cite{nak}
\begin{eqnarray}
\label{4a}
h_{add} =
\Omega \,m\, \omega_{00} \kappa\Bigg{[}
&2&\left(r^2 s_x - x \vec r \cdot \vec s\right)\\
&+&\mu \left(2 r^2 - \frac{\hbar}{m\omega_{00}}
(N+\frac{3}{2})\right)\,l_x \Bigg{]}.\nonumber
\end{eqnarray}

The two-body potential has the following structure
\begin{equation}
V = V_{PP} + V_{QQ} + V_{MM} + W_{\sigma \sigma}.
\label{2b}
\end{equation}
It includes a monopole pairing, 
$V_{PP} = - \sum_{\tau=p,n} G_{\tau} P_{\tau}^\dagger P_{\tau}$,
where $P_{\tau}^{\dagger} = \sum_{\chi} c_{\chi}^\dagger c_{\bar {\chi}}^{\dagger}$.
An index $\chi$ is labelling a complete set of the oscillator quantum numbers
($|\chi\rangle = |N l j m \rangle$) and the index ${\bar {\chi}}$ denotes the
time-conjugated state \cite{BM1}.
$V_{QQ}$ and $V_{MM}$ are, respectively,
separable quadrupole-quadrupole, 
$V_{QQ} = - \frac{1}{2} \sum_{T=0,1} \kappa (T)
\sum_{r=\pm} \sum_{\mu=0,1,2} (\tilde Q_\mu [^{T}_{r}])^2$, 
and monopole-monopole,
$V_{MM}= - \frac{1}{2} \sum_{T=0,1} \kappa (T) (\tilde M
[^{T}_{r=+}])^2$, potentials.
$V_{\sigma \sigma}$ is a spin-spin interaction,
$V_{\sigma \sigma} = - \frac{1}{2}
\sum_{T=0,1} \kappa_{\sigma} (T)\sum_{r=\pm} \sum_{\mu=0,1} (s_\mu [^{T}_{r}])^2$.
We recall that the K quantum number (a projection of the angular momentum
on the quantization axis) is not conserved
in rotating nonaxially deformed systems.
However, the cranking Hamiltonian (\ref{3}) adheres to
the $D_2$ spatial symmetry with respect to rotation
by the angle $\pi$ around the rotational axis $x_1$.
Consequently, all rotational states can be classified by the
quantum number called signature $r=\exp (-i\pi\alpha)$ leading
to selection rules for the total angular momentum
$I=\alpha + 2n$, $n=0,\pm 1, \pm 2 \ldots $
In particular, in even-even nuclei the yrast band
characterized by the positive signature quantum number
$r=+1 \,(\alpha = 0)$ consists of even spins only.
All the one-body fields have good z component of isospin operator $t_z$ and 
signature $r$.
Multipole and spin-multipole fields of good
signature are defined in Ref.\cite{JK}.
The tilde indicates that monopole and quadrupole fields are
expressed in terms of doubly stretched coordinates
$\tilde{x}_i=(\omega_i/\omega_0)\,x_i$ \cite{ds}.

Using the generalized Bogoliubov transformation for quasiparticles
(for example, for the positive signature quasiparticle we have
$\alpha_i^+=\sum_k{\cal U}_{ki} c_k^+ + {\cal V}_{\bar k i}c_{\bar k}$)
and  the variational principle (see details in Ref.\onlinecite{KN}), we obtain
the Hartree-Bogoliubov (HB) equations
for the positive signature quasiparticle energies $\varepsilon _i $ (protons or neutrons)
\begin{equation}
\label{hb}
\Biggl(
\begin{array}{cc}
h(1)& \triangle \\
\triangle &h(2)\\
\end{array}
\Biggr)
\Biggl(
\begin{array}{c}
{\cal U}_i\\
{\cal V}_i\\
\end{array}
\Biggr) =\varepsilon _i
\Biggl(
\begin{array}{c}
{\cal U}_i\\
{\cal V}_i\\
\end{array}
\Biggr).
\end{equation}
Here, $h(1)_{kl}=(\tilde H_0)_{kl}-\Omega (I_1)_{kl}$,
$(h(2))_{kl}=-(\tilde H_0)_{kl}-\Omega (I_1)_{kl}$ and 
$\Delta_{kl}=-\delta_{kl}G_{\tau} <P_{\tau}> $ and
 $|k>$ denotes a s.p. state of a Goodman spherical basis (see Ref.\onlinecite{JK}).
It is enough to solve the HB equations  for the positive signature, since
the negative signature eigenvalues and eigenvectors are
obtained from the positive ones according to relation
\begin {equation}
\left(-\varepsilon _i , {\cal U}_i, {\cal V}_i \right) 
\rightarrow \left( {\varepsilon} _{\tilde{i}},
{\cal V}_{\tilde{i}}, {\cal U}_{\tilde{i}} \right)
\end {equation}
where the state $\tilde{i}$ denotes the signature partner of $i$ .
For a given value of the rotational frequency $\Omega$
the quasiparticle (HB) vacuum state is defined as 
$\alpha_{i}|\rangle = \alpha_{\tilde{i}}|\rangle = 0$.

The solution of a system nonlinear  HB equations (\ref{hb})
is a nontrivial problem. In principle, the pairing gap should be determined
self-consistently at each rotational frequency.
However, in the vicinity of the backbending, the solution becomes highly unstable.
In order to avoid unwanted singularities for certain
values of $\Omega$, we followed
the phenomenological prescription \cite{wys}
\begin{equation}
\label{17}
\Delta_{\tau}(\Omega) \,=\,
\left \{
\begin{array}{l}
\Delta_{\tau}(0)\,[1-\frac{1}{2} (\frac{\Omega}{\Omega_c})^2 \,]
\qquad
\,\,\,\,\,\,\Omega < \Omega_c \\
\Delta_{\tau}(0)\,
\frac{1}{2} (\frac{\Omega_{c}}{\Omega})^2 \qquad \qquad \, \,
\,\,\,\,\,\,\,\,\Omega > \Omega_c, \\
\end{array},
\right.
\end{equation}
where $\Omega_c$ is  the critical rotational frequency
of the first band crossing.

In general, in standard calculations with the Nilsson or Woods-Saxon potentials, 
the equilibrium deformations are determined with the aid of 
the Strutinsky procedure \cite{Rag}. The procedure, being a very effective tool
for an analysis of experimental data related to the ground or yrast states, 
produces deformation parameters that are slightly different from those of the 
mean field calculations. The use of the former parameters (based on the 
Strutinsky procedure) for the RPA  violates the self-consistency between the
mean field and the RPA description. Therefore, to keep a self-consistency 
between the mean field and the RPA as much as possible, we use the recipe 
described below. 

It is well known \cite{ds,HN02,n2,A03} that for a deformed
HO Hamiltonian,
the quadrupole fields in double-stretched coordinates fulfill the stability conditions
\begin{equation}
\langle\tilde Q_\mu\rangle = 0, \qquad \mu=0,1,2
\label{new}
\end{equation}
if nuclear self-consistency
\begin{equation}
\omega_{1}^2 \langle x_{1}^2 \rangle = \omega_{2}^2 \langle x_{2}^2 \rangle=
\omega_{3}^2 \langle x_{3}^2\rangle
\end{equation}
is satisfied in addition to the volume conserving  constraint.
Here,  $<...>$ means the averaging over the mean field vacuum state of the rotating system.
In virtue of the stability conditions (\ref{new}),
the interaction will not distort further the deformed HO potential,
if the latter is generated  as a  Hartree field.  To this purpose, one starts
with an isotropic HO potential of frequency $\omega_0$ and, then,
generates the deformed part of the potential from the (unstretched) separable
quadrupole-quadrupole (QQ) interaction. The outcome of this procedure is
\begin{equation}
V_{HO} = \frac{m\omega_0^2 r^2}{2} - m\omega_0^2 \beta \,cos\gamma \,
Q_{0}[^{\,0}_{+}] - m\omega_0^2 \beta \,sin\gamma
Q_{2}[^{\,0}_{+}]
\end{equation}
where one can use the following parameterization of the quadrupole deformation in terms
of $\beta$ and $\gamma$ (see, for example, Ref.\onlinecite{RS}):
\begin{eqnarray}
m\omega_0^2 \beta \,cos\gamma &=&
 \kappa[0] \langle Q_{0} [^{\,0}_{+}]\rangle\nonumber\\
m\omega_0^2 \beta \,sin\gamma &= &-\kappa[0] \langle Q_{2} [^{\,0}_{+}]\rangle.
\label{Hartree}
\end{eqnarray}
The triaxial form given by Eq. (\ref{TriHO}) follows from
defining
\begin{equation}
\omega^{2}_{i} = \omega^{2}_{0}\left[ 1 - 2\beta \sqrt{\frac{5}{4\pi}}cos(\gamma
- \frac{2\pi}{3}i)\right], \,\, i = 1,2,3
\end{equation}
Here, we follow the convention on the sign of $\gamma$-deformation accepted
in Ref.\onlinecite{RS}.
The Hartree conditions have the form given by Eqs.(\ref{Hartree})
only for a spherical HO potential plus the QQ forces.
Quite often, Eqs.(\ref{Hartree}) are considered as self-consistent conditions
for pairing+QQ model interaction. In practice, the use of these conditions is based
on $\Delta N=0$ mixing in a small configuration space around the Fermi energy, which
includes only 3 shells. This restriction limits a description 
of physical observables like a mean field radius and vibrational excitations. 
Furthermore, if $\Delta N=2$ mixing is included,  the RPA correlations are 
overestimated \cite{A03}. In addition, the QQ-forces, without the volume conservation 
condition, fail to yield a minimum for a mean field energy of rotating superdeformed  
nuclei \cite{A03}. Due to all these facts, we allow small 
deviations from Eqs.(\ref{Hartree}) and enforce only the stability conditions
(\ref{new}), which are our self-consistent conditions for the mean field calculations. 
These, in fact, hold also in the presence of pairing
(see below)  and ensure
the separation of the pure rotational mode from the intrinsic
excitations for a cranked harmonic oscillator \cite{n2}.

\subsection{Some HB results}

As was mentioned in Introduction, for our calculations we choose
$^{156}Dy$ and $^{162}Yb$.
There are enough available experimental data on spectral characteristics
and electromagnetic decay of high spins in these nuclei \cite{nndc}.
It is also known that these nuclei possess axially symmetric ground states and 
exhibit the backbending behaviour at high spins. Moreover, nuclei with Z$\sim 66$ and 
N$\sim 90$ attract a theoretical attention for
a long time, since the cranking model predicts that high-j quasiparticles drive 
rotating nuclei to triaxial shape \cite{fr90}. 
\begin{figure}[ht]
\includegraphics[height = 0.2\textheight]{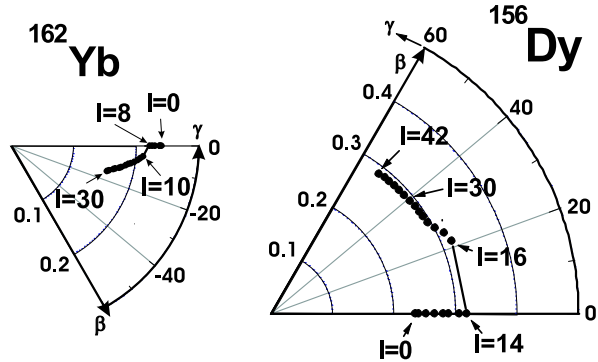}
%\centerline{\psfig{figure=fig1.eps,width=3.2in,clip=}}
\caption{
Equilibrium deformations in $\beta$-$\gamma$ plane
as a function of the angular momentum $I = \langle {\hat I}_1 \rangle-1/2$
(in units of $\hbar$).}
\label{eqd}
\end{figure}

We perform the following calculations:
\begin{enumerate}
\item
The Hamiltonian (\ref{3}) includes the term $\hat h_{add}$ (\ref{4a}) that 
restores the local Galilean invariance broken in the Nilsson potential. 
This calculation will be further refereed as the calculation \textbf{I}.
\item The Hamiltonian (\ref{3}) does not include the term $\hat h_{add}$. 
Since this term is responsible for the correct behaviour
of the moment of inertia, such calculations provide the answer about its 
importance, for example, for the value of the band crossing frequency. 
All microscopic results reported in literature 
(excluding the analysis of octupole excitations \cite{nak}) do not 
include such term in the Nilsson potential. We will refer this calculation as the
calculation \textbf{II}
\end{enumerate}

Parameters of the Nilsson potential were taken from Ref.\cite{31}. These parameters
were determined from a systematic analysis of the experimental s.p.
levels of deformed nuclei of rare earth and actinide regions. In our calculations
we include all shells up to $N=9$ and this configuration space was sufficient
for the fulfilling $99\%$ of the sum rule for $E2$ strength 
(see Ref.\onlinecite{JK2}).
In contrast to the analysis of Refs.\onlinecite{mat1,mat2,shi1}, based on "single stretched" 
coordinate method that involves the $\Delta N = 2$ mixing only approximative,
we take into account the $\Delta N = 2$ mixing exactly.
This improves the accuracy of the mean field calculations.
For the values of the pairing gaps $\Delta_{\tau}(0)$ at zero
rotational frequency we use the results of Ref.\onlinecite{BRM}:
$\Delta_{N}(0) = 0.940$ MeV, $\Delta_{P}(0) = 0.985$ MeV for $^{156}$Dy and
$\Delta_{N}(0) = 0.967$ MeV, $\Delta_{P}(0) = 0.975$ MeV for $^{162}$Yb.
These values have been obtained in order to reproduce nuclear masses.
For quantization of angular momentum we use the equation
$<I_1>=I+1/2$. Here, the term $1/2$ is due to the Nambu-Goldstone mode
that appears in the RPA (see Ref.\onlinecite{A03}).

\begin{figure}[ht]
\includegraphics[height = 0.3\textheight]{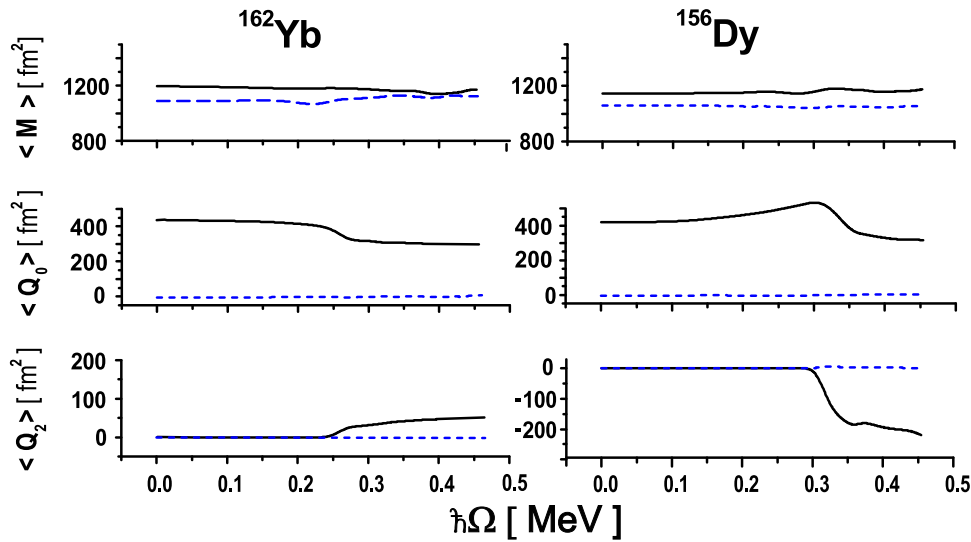}
%\centerline{\psfig{figure=fig2.eps,width=3.2in,height=0.15,clip=}}
\caption{
(Color online) The rotational behaviour of the calculated monopole and quadrupole moments.
The "double stretched"  and standard values are connected by dashed and solid line,
respectively.}
\label{qmom}
\end{figure}

As shown in Fig.\ref{eqd}, the triaxiality of the mean field sets 
in at the rotational frequency $\Omega_c$ 
which triggers a backbending in the considered nuclei.
We obtain the critical rotational frequencies, at which the first band crossing 
occurs, $\hbar \Omega_{c}=0.250$, $0.301$ MeV for $^{162}$Yb, $^{156}$Dy, respectively. 
The parameters so determined yield results in a better agreement with experiments, 
compared to the ones obtained in Ref.\cite{fr90} for $N\sim 90$.
Moreover, our equilibrium deformations are short from being
the self-consistent solutions of the HB equations. The doubly stretched 
quadrupole moments $ \langle{\tilde Q}_{0} [^{\,0}_{+}]\rangle$ and
$\langle{\tilde Q}_{2} [^{\,0}_{+}]\rangle$  are approximately
zero for all values of the equilibrium deformation parameters,
consistently with the stability conditions (\ref{new}) (see Fig.\ref{qmom}).
Indeed,  any deviation from the equilibrium values of the deformation parameters
$\beta$ and $\gamma$ results into a higher HB energy. 
The "double stretched" monopole moment $\langle M_{0} [^{\,0}_{+}]\rangle$ is
not far from the corresponding standard one.  The both values, the standard and the
"double stretched" monopole moments, are almost independent on $\Omega$.
We infer from the just discussed tests  that our solutions are close to the
self-consistent  HB ones. In contrast with our results,
in Ref.\onlinecite{fr90} the fixed parameters (deformation, pairing gap) 
were used for all values of the rotational frequency. 
In addition, the analysis of 
Ref.\onlinecite{fr90} was based on the fitted moments of inertia, that were kept 
constant for all rotational frequencies. It is evident that such 
an analysis can be only used for a discussion of a general trend. 

\begin{figure}[ht]
\includegraphics[height = 0.45\textheight]{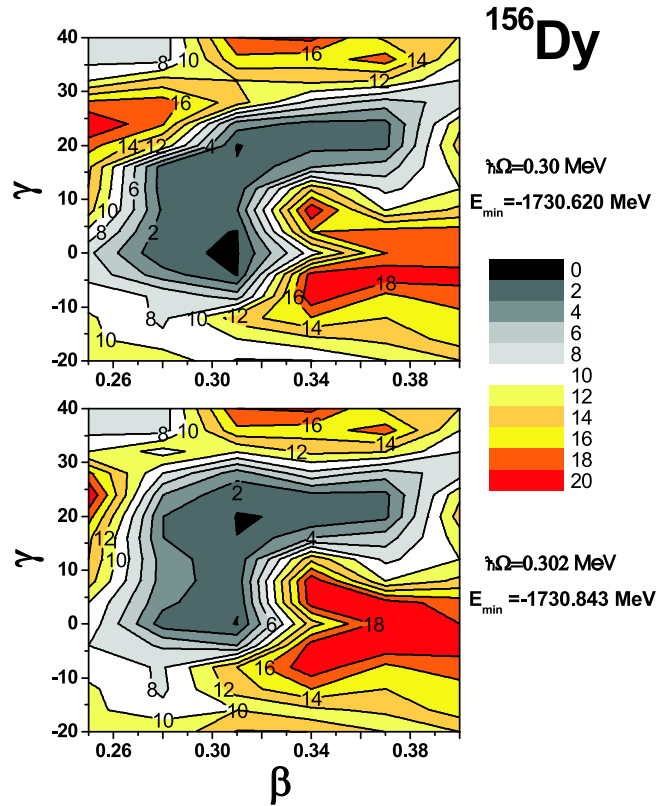}
%\centerline{\psfig{figure=fig3.eps,width=3.2in,clip=}}
\caption{
(Color online) The potential energy surface $E_{\Omega}(\beta ,\gamma)$
for $^{156}$Dy  before ($\hbar\Omega=0.300$ MeV) and after ($\hbar\Omega=0.302$ MeV) 
the transition point. There are two local minima with almost the 
same $\beta$ deformation and completely different $\gamma$-deformation.}
\label{pes1}
\end{figure}

We get more insight into the backbending mechanism if consider the potential energy 
surface near the transition point. As is shown in Fig.\ref{pes1}, the potential 
energy surface of the total mean field energy 
$E_{\Omega}(\beta ,\gamma) =\langle H_{\Omega}\rangle$, for
$^{156}$Dy  at $\hbar\Omega =0.300$ MeV  (before the bifurcation point) exhibits
the minimum for the axially symmetric shape which is 
lower than the minimum for strongly triaxial shape with $\gamma \approx 20^0$.
The increase of the rotational frequency breaks the axial symmetry and
a nucleus settles at the nonaxial minimum at $\hbar\Omega=0.302$ MeV 
(after the bifurcation point). Notice, the main difference in these minima is a 
strong nonaxial deformation of  the one minimum in contrast to the other, 
while the $\beta$- deformation is almost the same.

\begin{figure}[ht]
%\centerline{\psfig{figure=fig4.eps,width=2.2in,clip=}}
\includegraphics[height = 0.25\textheight]{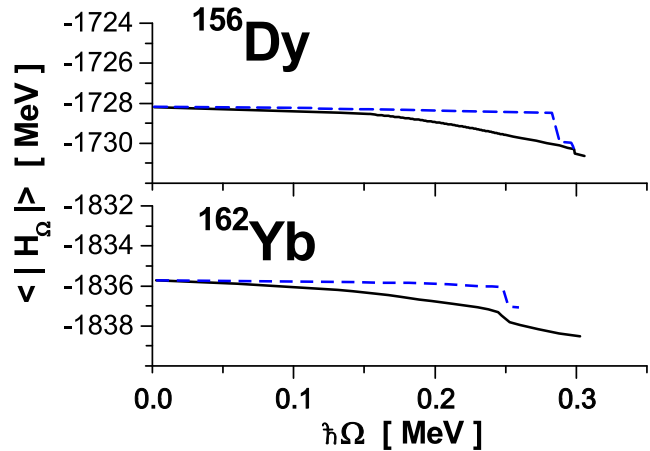}
\caption{
(Color online) The rotational dependence of the total mean field energy 
$E_{\Omega}(\beta, \gamma=0)=\langle H_{\Omega} \rangle$ for the
axially symmetric equilibrium deformation. Results  for  collective ($x_1$ axis)
and noncollective ($x_3$ axis) rotation are connected by solid and dashed line,
respectively. Results for $^{156}$Dy and $^{162}$Yb are displayed on
upper and  lower panel, respectively.}
\label{pes}
\end{figure}

Dealing with transitional nuclei, however,
the minimum becomes very shallow
for a collective (around  $x_{1}$ rotational axis) and non-collective
(around $x_3$ symmetry axis) rotation as the rotational frequency increases.
In fact, the energy minima for the collective 
($\approx \,-1730.6$ MeV)
and non-collective ($ \approx \,-1730.62$ MeV)
rotations are almost degenerate near the crossing point
of the ground with  the $\gamma$- band for  $^{156}$Dy (see
the upper panel of the Fig. \ref{pes}).
The energy difference is about 15 keV near the critical rotational frequency
where the backbending occurs.
At the transition point, the competition between
collective and non-collective rotations breaks the
axial symmetry and  yields nonaxial shapes.
{\it Does this behaviour correspond to a phase transition ?}

First, notice that a half of an experimental value for $\gamma$-vibrations
$\hbar \Omega_{K=2}/2= 0.414$ MeV (at $\hbar \Omega=0$) \cite{fot} 
is close to the collective rotational frequency $\hbar \Omega_c = 0.301$ MeV
at which the shape transition occurs.
Second, let us consider an
axially deformed system, defined by the Hamiltonian ${\tilde H}$
in the laboratory frame, that rotates about a symmetry axis z with
a rotational frequency $\Omega$. The angular momentum is a good
quantum number and, consequently,
$[\hat{J}_z,O_K^{\dagger}]=KO_K^{\dagger }$. Here,  the
phonon  $O_K^{\dagger}$ describes the vibrational state with $K$
being the value of the angular momentum carried by the phonons
$O_K^{\dagger}$ along the symmetry axis, $z$ axis. Thus, one
obtains 
\begin{equation} 
[H_{{\Omega}},O_K^{\dagger}]= [{\tilde H} -\Omega
\hat{J}_z, O_K^{\dagger }]= ({\tilde \omega}_K- K\Omega)
O_K^{\dagger } \equiv \omega_K O_K^{\dagger },
\label{man} 
\end{equation}
where ${\tilde \omega}_K$ is the phonon energy of the mode K in the laboratory
frame at $\Omega=0$.
This equation implies that at the rotational frequency
$\Omega_{cr}={\tilde \omega}_K/K$ one of the RPA frequency
$\omega_K$ vanishes in the rotating frame (see discussion in
Refs.\onlinecite{MN93,HN02,M96}). At this point of bifurcation we could expect the
SSB effect of the rotating mean field due to the appearance of the
Goldstone boson related to the multipole-multipole forces with
quantum number $K$. For an axially deformed system,
one obtains the breaking of the axial symmetry, since the lowest
critical frequency corresponds to $\gamma$-vibrations with $K=2$
\cite{MN93,HN02}. 
The rotation around collective $x_1$ axis couples, however, vibrational
modes with different $K$ and the critical rotational frequency is influenced 
by this coupling: it becomes lower. The value of $\Omega_{cr}$ can be affected by the 
degree of the collectivity of the vibrational excitations, as we will
see below. The most important outcome from this consideration is that in the vicinity 
of the shape transition there is {\it an anomalously low vibrational mode related to 
the deformation parameter $\gamma$}.

\begin{figure}[ht]
\includegraphics[height = 0.45\textheight]{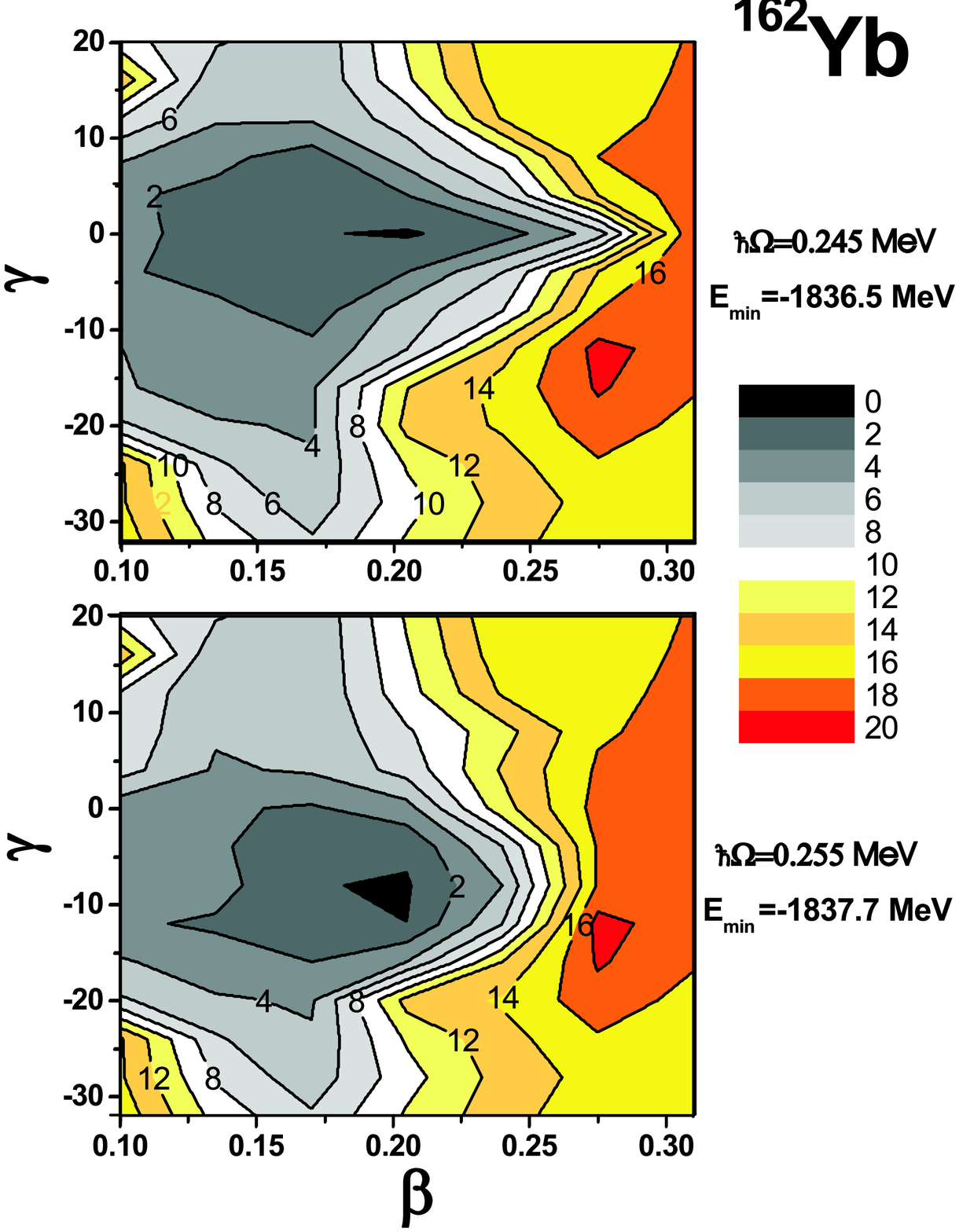}
%\centerline{\psfig{figure=fig5.eps,width=3.2in,clip=}}
\caption{
(Color online) The potential energy surface $E_{\Omega}(\beta ,\gamma)$
for $^{162}$Yb  before ($\hbar\Omega=0.245$ MeV) and after ($\hbar\Omega=0.255$ MeV) 
the transition point.
}
\label{pes2}
\end{figure}

For $^{162}Yb$ the shape transition takes place at the
rotational frequency  $\hbar\Omega_c \approx 0.25$ MeV, while the experimental 
bifurcation point (a half of the $\gamma$-vibrational excitation energy 
at $\hbar \Omega=0$) is $\hbar \Omega_{\it cr} \approx 0.45$ MeV. 
The energy difference at the transition point
between the collective  ($\approx \,-1837.7$ MeV) and the non-collective
($\approx \,-1836.6$ MeV) minima is still large $\sim 1$ MeV (see the lower panel
of Fig.\ref{pes}). As one can see in Fig.\ref{pes2},
the difference between the axially symmetric minimum at 
$\hbar\Omega=0.245$ MeV (before the transition) and the nonaxial 
one at $\hbar\Omega=0.255$ MeV (after the transition) for a collective rotation is 
about $\sim 1.2$ MeV. The deformation parameters change smoothly  with the
increase of the rotational frequency at the transition point (see Fig.\ref{eqd}).
It appears, that for $^{162}Yb$ there is a different 
mechanism responsible for the observed backbending.

To elucidate the different character of the shape 
transition from axially symmetric to the triaxial shape and its relation to 
a phase transition, we consider 
potential landscape sections in the vicinity of the shape transition. 
The phase transition is detected by means of the order parameter
as a function of a control parameter \cite{Lan}. In our case, the deformation
parameters $\beta$ and $\gamma$ are natural order parameters,
while the rotational frequency $\Omega$ is a control parameter 
that characterizes a rotational state in the rotating frame.
Since we analyze a shape transition from the axially symmetric shape
($\gamma = 0$) to the triaxial one ($\gamma \neq 0$), we choose only
the deformation parameter $\gamma$ as the order parameter 
that reflects the broken axial symmetry.
Such a choice is well justified, since the deformation parameter $\beta$ 
preserves its value before and after the shape transition in both nuclei: 
$\beta_t \approx 0.2$ for $^{162}Yb$ and 
$\beta_t \approx 0.31$ for $^{156}Dy$. Thus, we consider 
a mean field value of the cranking Hamiltonian,
$E_{\Omega}(\gamma;\beta_t) \equiv \langle H_{\Omega} \rangle$, for different 
values of $\Omega$ (our state variable) and $\gamma$ (order parameter) 
at fixed value of $\beta_t$ . 

\begin{figure}[ht]
\includegraphics[height = 0.35\textheight]{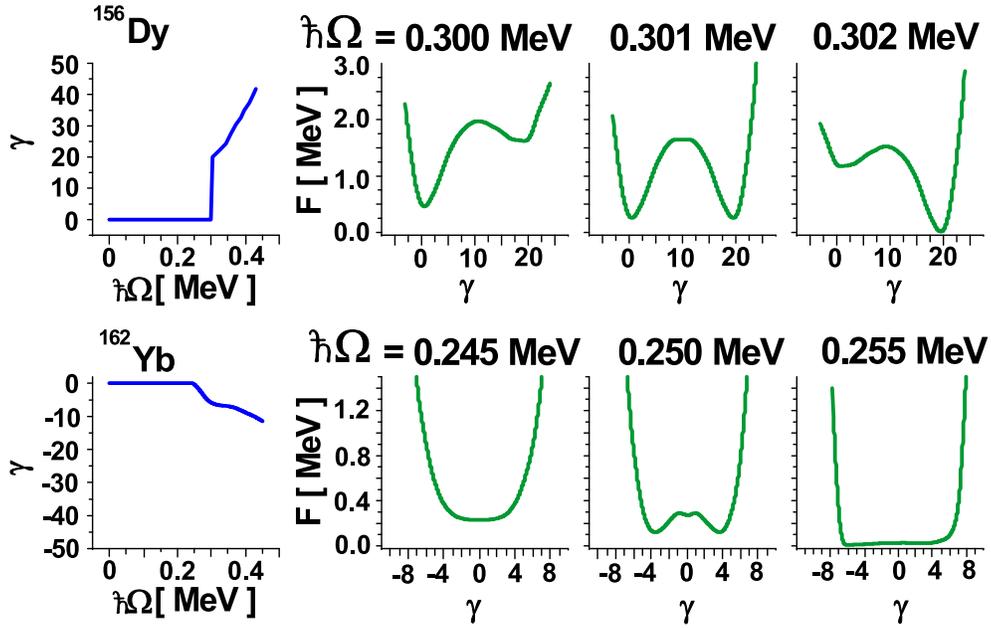}
%\centerline{\psfig{figure=fig6.eps,width=3.6in,clip=}}
\caption{
(Color online) 
The rotational dependence of the order parameter $\gamma$ and
the energy surfaces sections  $F(\Omega,\gamma)=E(\gamma,\beta_t)-E_{min}$ 
for $^{156}Dy$ (top) and $^{162}Yb$ (bottom) before and after the transition point. 
The energy is given relative  to the value $E_{min}=E_{\Omega}(\beta_t,\gamma)$ at 
 $\hbar\Omega=0.255,\,\,0.302$ MeV for $^{162}Yb$, $^{156}Dy$,
respectively.
}
\label{phatr}
\end{figure}

For $^{156}Dy$ we observe the emergence of the order parameter $\gamma$ 
above the critical value $\hbar\Omega_c=0.301$ MeV 
of the control parameter $\Omega$ (see a top panel in Fig.\ref{phatr}).
Below and above the transition point there is a unique phase  whose properties
are continuously connected to one of the coexistent phases at the transition point.
The order parameter changes discontinuously as the nucleus passes through the
critical point from axially symmetric shape to the triaxial one. 
The polynomial fit of the potential landscape section 
at $\hbar\Omega_c=0.301$ MeV yields the following expression
\begin{equation}
F(\Omega;\gamma) =F_0(\Omega)+ F_2(\Omega) \gamma^2 - F_3(\Omega) \gamma^3 + 
F_4(\Omega) \gamma^4 ,
\end{equation} 
where the coefficients $F_0(\Omega)=0.3169$ MeV, $\gamma$ in degrees and 
$F_2(\Omega)=0.12239$, $F_3(\Omega)=0.009199$,
$F_4(\Omega)=1.7\times10^{-4}$ are defined in  corresponding units.
We can transform this polynomial to the form
\begin{equation}
{\bar F}=\frac{F(\Omega;\gamma) -F_0(\Omega)}{\bar {F_0}}
\approx \alpha \frac{\eta^2}{2}- \frac{\eta^3}{3} +\frac{\eta^4}{4}
\label{gen}
\end{equation}
where
\begin{equation}
{\bar {F_0}}=\frac{(3F_3)^4}{(4F_4)^3},\qquad 
\alpha=\frac{8F_2F_4}{9F_3^2}, \qquad \eta=\frac{4F_4}{3F_3}\gamma
\end{equation}
The expression (\ref{gen}) represents the generic form of  the anharmonic model 
of {\it the structural first order phase 
transitions} in condensed matter physics (cf \cite{krum}). 
The condition $\partial {\bar F}/\partial \eta=0$ determines the following 
solutions for the order parameter $\eta$
\begin{equation}
\eta=0,\qquad \eta=\frac{1\pm\sqrt{1-4\alpha}}{2}
\end{equation}
If $\alpha>1/4$, the functional ${\bar F}$ has a single minimum at $\eta=0$.
Depending on  values of $\alpha$, defined in the interval
$0<\alpha<1/4$, the functional ${\bar F}$ manifests the transition
from one stable minimum at zero order parameter via one minimum+metastable state
to the other stable minimum with the nonzero order parameter.
In particular, at the universal value of $\alpha=2/9$ the functional
${\bar F}$ has two minimum values with 
${\bar F}=0$ at $\eta=0\rightarrow \gamma\approx0^0$ and 
$\eta=2/3\rightarrow \gamma \approx27^0$
and a maximum at $\eta=1/3\rightarrow \gamma\approx13.5^0$. The correspondence 
between the actual value $\gamma\approx 20^0$ and the one obtained from the 
generic model  is quite good. Thus the backbending in $^{156}$Dy possesses typical 
features of {\it the first order phase transition}. 

In the case of $^{162}Yb$ the energy $E(\Omega;\gamma)$  and 
the order parameter (Fig.\ref{phatr}) are smooth  functions in the vicinity of 
the transition point $\Omega_c$. This implies that two phases, 
$\gamma=0$ and $\gamma\neq0$, on either side of the transition point should 
coincide. Therefore, for $\Omega$ near the transition point $\Omega_c$
we can expand our functional 
$F(\Omega,\gamma)=E_{\Omega}(\gamma,\beta_t)-E_{min}$ (see Fig.\ref{phatr}) 
in the form
\begin{equation}
F(\Omega;\gamma) = F_1(\Omega) \gamma + 
F_2(\Omega) \gamma^2 + F_3(\Omega) \gamma^3 + F_4(\Omega) \gamma^4 + \ldots
\label{5}
\end{equation} 
The conditions of the phase equilibrium  (further we 
restrict the expansion (\ref{5}) up to the terms with $\gamma^4$)
\begin{equation}
\frac{\partial F}{\partial \gamma} = F_1(\Omega) + F_2(\Omega) 2 \gamma + 
F_3(\Omega) 3 \gamma^2 + F_4(\Omega) 4 \gamma^3
= 0
\label{lmin1}
\end{equation} 
\begin{equation}
\frac{\partial^2 F}{\partial \gamma^2} = 2 F_2(\Omega) + 6 F_3(\Omega) \gamma + 
12 F_4(\Omega) \gamma^2 \geq 0 \,,
\label{lmin2}
\end{equation} 
which should be valid for all values of $\Omega$ and $\gamma$ (including 
$\gamma=0$), yield
\begin{equation}
F_1(\Omega) = 0
\label{lmin3}
\end{equation}
Eqs. (\ref{lmin1}) and (\ref{lmin2}) can be rewritten as
\begin{equation}
2 F_2(\Omega) \gamma + 3 F_3(\Omega) \gamma^2 + 4 F_4(\Omega) \gamma^3 = 0
\label{lmin4}
\end{equation}
\begin{equation}
2 F_2(\Omega) + 6 F_3(\Omega) \gamma + 12 F_4(\Omega) \gamma^2 \geq 0
\label{lmin5}
\end{equation}
which implies the following inequality
\begin{equation}
2 F_4(\Omega) \gamma^2 \geq  F_2(\Omega)
\label{lmin6}
\end{equation}
This inequality holds for all values of $\gamma$ (including 
$\gamma=0$ at $\Omega=\Omega_c$), which leads to $F_2(\Omega = \Omega_c) \leq 0$.
On the other hand, from the stability condition Eq.(\ref{lmin2}) 
at the transition point $\Omega_c$ and $\gamma=0$, we also have
$F_2(\Omega = \Omega_c) \geq 0$. The both inequalities can coincide only
when $F_2(\Omega = \Omega_c) = 0$.
Using the result $F_1(\Omega_c)=F_2(\Omega_c)=0$ and the fact that
all phases at the transition point should coincide, we obtain from Eq.(\ref{lmin1})
that $F_3(\Omega = \Omega_c) = 0$.
Assuming that $F_3=0$ for all $\Omega$, the minimum condition
 Eq.(\ref{lmin4}) yields
the following solution for the order parameter
\begin{equation}
\gamma_1=0 \,, \quad
\gamma_{2,3}^2 = - \frac{F_2(\Omega)}{2\,F_4(\Omega)} = 
\left\{
\begin{array}{ll}
\neq 0 & for \,\, \Omega\neq \Omega_c \\
= 0    & for \,\, \Omega = \Omega_c
\end{array}
\right.
\label{par2}
\end{equation}
Since at the transition point $F_2(\Omega_c)=0$, one can propose 
the following definition of the function $F_2(\Omega)$: 
\begin{equation}
F_2(\Omega) \approx \frac{dF_2(\Omega)}{d\Omega}\, \left(\Omega - \Omega_c \right)
\label{par3}
\end{equation}
Thus,  we have $\gamma\sim (\Omega-\Omega_c)^\nu$ and the critical exponent 
$\nu=1/2$, in accord with the classical Landau theory, where the temperature is 
replaced by the rotational frequency.

Our extension of the Landau-type approach for rotating nuclei  
is nicely confirmed by the numerical results.
The polynomial fit of the energy potential surfaces for $^{162}Yb$ (Fig.\ref{phatr})
yields $F_1(\Omega)=F_3(\Omega)=0$ for all considered values of the rotational 
frequencies and $F_2(\Omega_c)=0$ at $\hbar \Omega_c = 0.25$ MeV. 
Moreover, in the vicinity of $\Omega_c$ we obtain $dF_2(\Omega)/\hbar d\Omega 
\approx -3.5$ ($F_4(\Omega)>0$ for all $\Omega$). In an agreement with 
Eqs.(\ref{par2}),(\ref{par3}), we have only the phase 
$\gamma =0$ for $\hbar \Omega < \hbar \Omega_c $ 
and the phase $\gamma \neq 0$ for $\hbar \Omega > \hbar \Omega_c $.  
The energy surfaces are symmetric with regard of
the sign of $\gamma$ and this also supports the idea that the effective energy $F$ 
can be expressed as an analytic function of the order parameter $\gamma$. 
Thus, the backbending in  $^{162}Yb$ can be classified as {\it the phase transition 
of the second order}.

\subsection{Quasiparticle spectra}

\begin{figure}[th]
\includegraphics[height = 0.4\textheight]{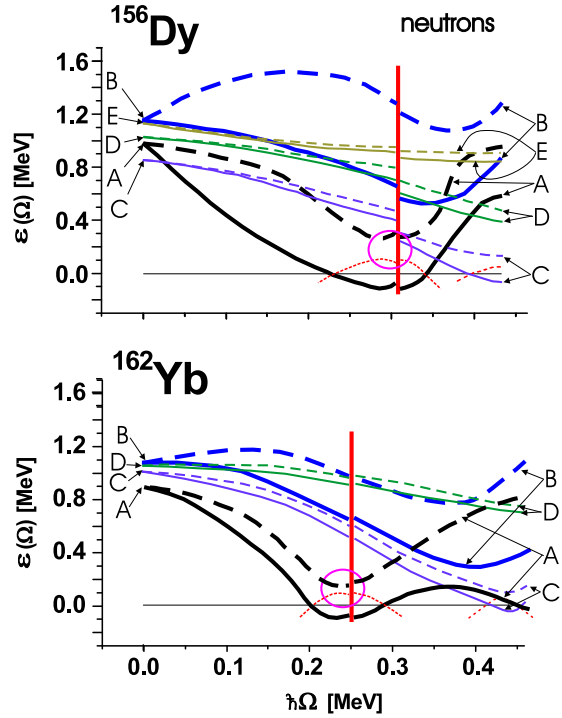}
%\centerline{\psfig{figure=fig7.eps,width=3.2in,clip=}}
\caption{
(Color online) Lowest quasineutron energies for $^{156}$Dy (upper panel) 
and $^{162}$Yb (lower panel). Thick (thin)  lines are used for the positive 
(negative) parity states. The positive (negative) signature states 
are connected by solid (dashed)  lines. At $\Omega =0$ the levels A, B,C, D, E 
correspond to the Nilsson states: 
3/2[651] (the subshell $i_{13/2}$ ), 1/2[660]) (the subshell $i_{13/2}$), 
3/2[521] (the subshell $h_{9/2}$), 5/2[521](the subshell $f_{7/2}$), 
11/2[505] (the subshell $h_{11/2}$), respectively. 
The shape transition point is denoted by the vertical line.
The quasicrossing is surrounded by a circle.
}
\label{qrn}
\end{figure}

\begin{figure}[th]
\includegraphics[height = 0.4\textheight]{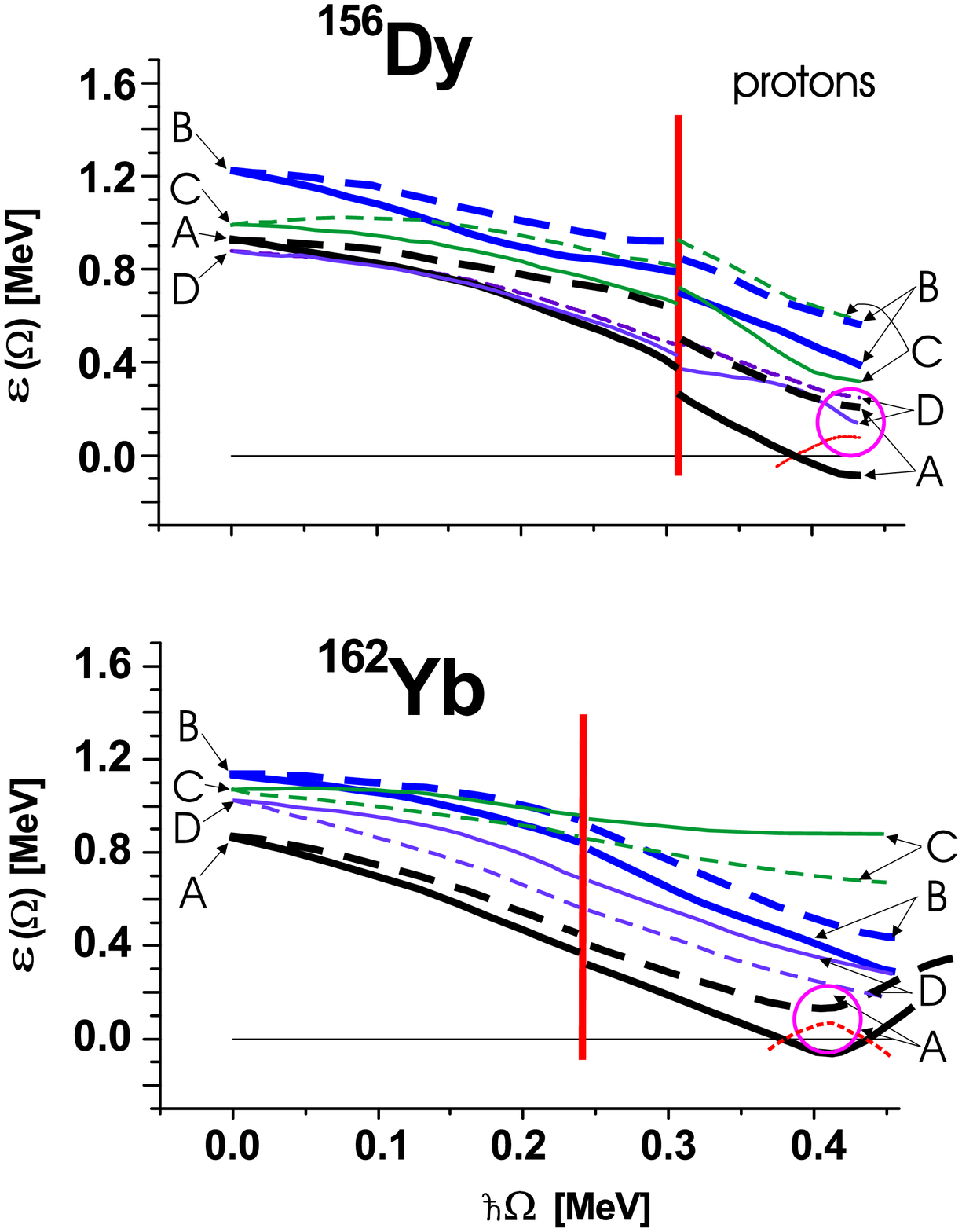}
%\centerline{\psfig{figure=fig8.eps,width=3.2in,clip=}}
\caption{
(Color online) Lowest quasiproton energies for $^{156}$Dy (upper panel) and 
$^{162}$Yb (lower panel). 
Thick (thin) lines are used  for the negative (positive) parity states. 
At $\Omega =0$ the levels A,B,C,D  correspond to the Nilsson states: 
7/2[523]  (the subshell $h_{11/2}$), 5/2[532] (the subshell $h_{11/2}$), 
3/2[411](the subshell $d_{5/2}$), 5/2[413] (the subshell $g_{7/2}$), 
respectively, in $^{156}$Dy;  7/2[523] (the subshell $h_{11/2}$), 
9/2[514] (the subshell $h_{11/2}$), 5/2[402] (the subshell $d_{5/2}$), 
7/2[404]  (the subshell $g_{7/2}$), 
respectively, in $^{162}$Yb. 
The shape transition point is denoted by the vertical line.
The quasicrossing is surrounded by a circle.
}
\label{qrp}
\end{figure}

To understand the microscopic origin of the quantum shape-phase transitions,
we analyse first the quasiparticle spectra and the rotational evolution of 
different observables like quadrupole moments and moments of inertia.
A numerical analysis of the expectation value of the nonaxial quadrupole moment
\begin{equation}
\langle \hat Q_{2}[^{\,0}_{+}] \rangle \,=\, \sum_{kl} 
{\langle} k| \hat Q_{2}[^{\,0}_{+}]|l {\rangle}\,
\sum_{i} \left[ {\cal V}_{i\bar k} {\cal V}_{i\bar l} \,+
\, {\cal V}_{\bar i k} {\cal V}_{\bar i l}\right]
\end{equation}
shows that in $^{162}$Yb  high-j neutron and proton orbitals that 
belong to $i_{13/2}$ and $h_{11/2}$ subshells, respectively, 
give the main contribution to the the expectation value. 
The nonaxial deformation grows due to the rotational alignment (RAL).
The crossing frequencies, where the configurations with aligned quasiparticles 
become yrast, are order of $\Omega_c\approx 2\Delta/i$, where $i$ is an aligned 
angular momentum carried by quasiparticles. Since the 
neutron gap is smaller than the proton gap, one may expect that the 
backbending should occur due to the alignment of the quasineutron orbitals
that could contribute $i\sim 8\hbar$. 

We trace the rotational evolution of quasiparticle orbitals in the rotating 
frame (routhians) as a function of the equilibrium parameters 
($\varepsilon$, $\gamma$, $\Delta$). 
At $\Omega=0$ each orbital is characterized by the asymptotic Nilsson quantum numbers.
However, these numbers lose their validity in the rotating case due to a strong mixing. 
Hereafter, they are used only for convenience.
The analysis of the routhians for neutrons (Fig.\ref{qrn}) and 
for protons (Fig.\ref{qrp}) indicates that the lowest quasicrossings 
occur: at $\hbar\Omega\approx0.275(0.42)$ MeV for neutron (proton) system 
in $^{156}$Dy; at $\hbar\Omega\approx0.245(0.41)$ MeV for neutron (proton) system 
in $^{162}$Yb. We recall that the shape-phase transition occurs at 
$\hbar\Omega_c\approx 0.25, 0.3$ MeV in $^{162}$Yb, $^{156}$Dy, respectively. 
The proximity of the critical point to the two-quasiparticle neutron 
quasicrossing in both investigated nuclei (especially, in $^{162}$Yb)  
suggests that the alignment of a pair $i_{13/2}$ is the main mechanism that 
drives both nuclei to triaxial shapes (cf \cite{fr90,scr}).
This mechanism itself, however, does not provide the explanation for the different 
character of the shape-phase transition. 
We recall that the routhians exhibit the dynamics of {\it noninteracting 
quasiparticles}. Evidently, the interaction between quasiparticle orbitals is 
important and could change substantially as a function of the neutron and 
proton numbers. Indeed, as we will see below (see Sec.III), this explains the 
type of shape-phase transitions discussed in the present section.

\subsection{Inertial properties}

\begin{figure}[ht]
%\centerline{\psfig{figure=fig9.eps,width=3.2in,clip=}}
\includegraphics[height = 0.45\textheight]{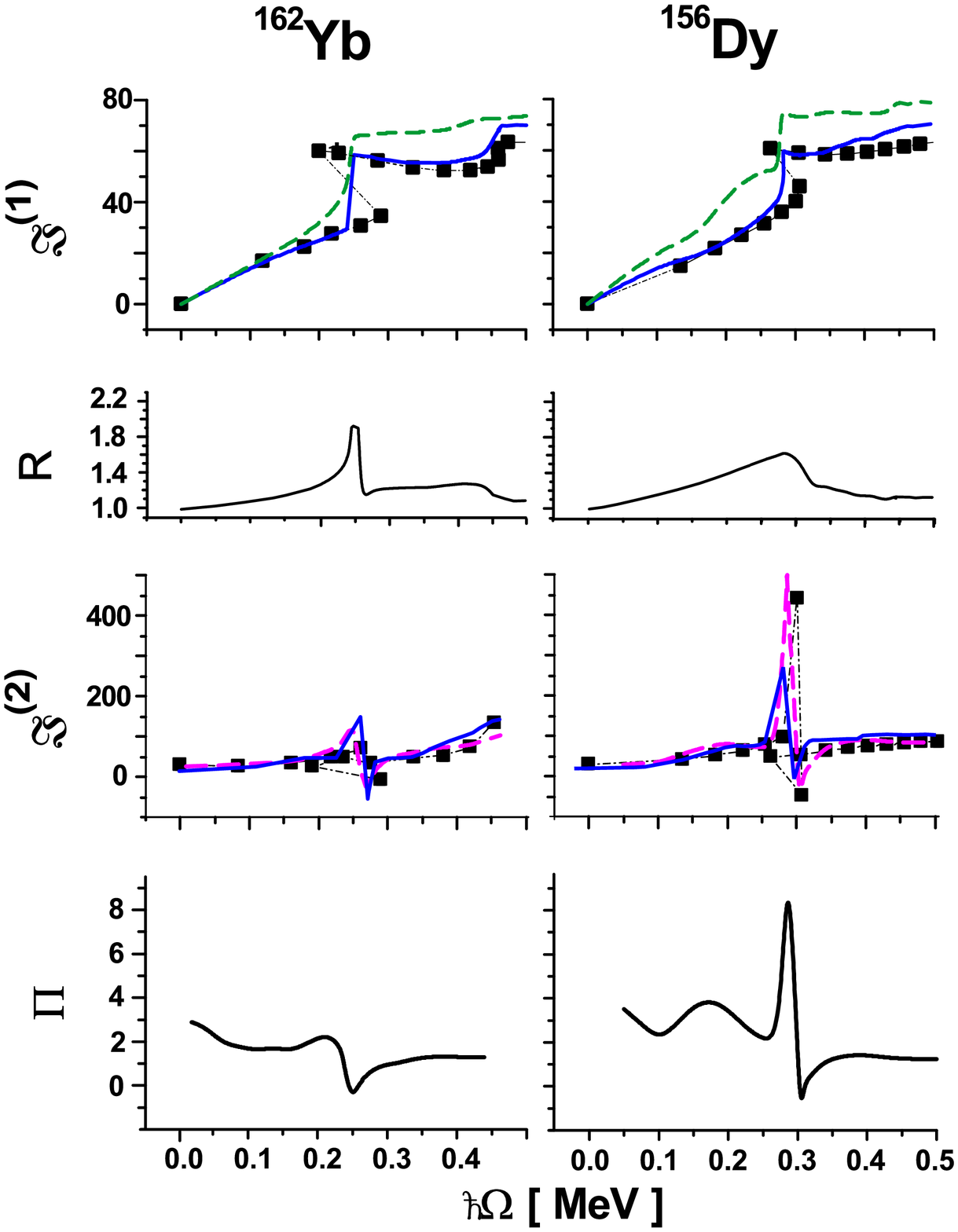}
\caption{
(Color online) The kinematical $\Im^{(1)}_{1}(\Omega) $ (top panels) and dynamical
$\Im^{(2)}_{1}(\Omega)$ (second panel from the bottom)
moments of inertia (in units $\hbar^2/MeV)$ are compared with the corresponding 
experimental values (filled squares). Experimental values are connected by 
dash-dotted line to guide eyes.
The results for the kinematical moment of inertia 
$\Im^{(1)}_{1}(\Omega) $, with (the calculations \textbf{I}) and without 
(the calculations \textbf{II}) the additional term $\hat h_{add}$, Eq.(\ref{4a}), 
are connected by solid and dashed lines,
respectively. The second panel (from the top)  displays the ratio
$R=\Im^{(1)}_{II}/\Im^{(1)}_{I}$ for each nucleus.
The experimental and calculated dynamical moment of inertia $\Im^{(2)}_1 $ (solid line)
(obtained by the calculation \textbf{I}) are compared with
the Thouless-Valatin inertia moment $\Im_{RPA}$ , Eq.(\ref{TV}) (dashed line).
In the bottom panels the ratio, $\Pi=\Im^{(2)}_1/\Im^{(1)}_1$
is displayed.
}
\label{moi}
\end{figure}

The moment of inertia is the benchmark for microscopic models of
nuclear rotation. To understand the interplay between s.p. degrees of freedom and
collective effects we calculate the kinematical
$\Im^{(1)}_{1} = \langle \hat I_{1} \rangle /\Omega$ and dynamical
$\Im^{(2)}_{1}= d\langle \hat I_{1} \rangle /d\Omega$
moments of inertia and compare with the corresponding experimental values.
The kinematical moment of inertia $\Im^{(1)}_{1}$ reflects the
collective properties of the rotating mean field. The dynamical moment of inertia 
due to obvious relation $\Im^{(2)}_{1} = \Im^{(1)}_{1}+\Omega d\Im^{(1)}_{1}/d\Omega$
is very sensitive to structural changes of the mean field. It
reflects the rearrangement of  {\it the two-body interaction} upon rotation
which leads to level crossings and shift in the deformation.
In fact, the dynamical moment of inertia provides a definite criteria on the
self-consistency of the rotating mean field when it is compared with the 
Thouless-Valatin moment of inertia calculated in the RPA. 
A full self-consistency is achieved if they are equal (see discussion in
Ref. \cite{n2}).

In Fig.\ref{moi} the experimental and theoretical values of kinematical (upper
panels) and dynamical (lower panels) moments of inertia are compared for
$^{162}$Yb (left panels) and $^{156}$Dy (right panels).
We remind that the calculations \textbf{I} (\textbf{II}) 
include (not include) the term $\hat h_{add}$, Eq.(\ref{4a}).
While both calculations reproduce the rotational evolution
of the kinematical moment of inertia, the agreement with the experimental data is
much better for calculations \textbf{I}. It is interesting to note that with the 
increase of the rotational frequency  the ratio $\Im^{(1)}_{I}/\Im^{(1)}_{II}$ 
increases from $\sim 50\%$ at $\hbar \Omega \sim 0.2$ MeV
to $\sim 85\%$ at $\hbar \Omega \sim 0.5$ MeV. 
Partially, the effect of broken Galilean symmetry is reduced
due to the alignment of the high-j intruder states with a large orbital momentum $l$.
These states contribute to the collective angular momentum and decrease the effect
of the $\vec l\,^2$ term in the Nilsson potential.

Although one observes a similar pattern for the backbending 
in the considered nuclei (upper panels, Fig.\ref{moi}), a different response of 
a nuclear field upon the rotation becomes more evident with the aid of 
the experimental dynamical moment of inertia 
$\Im^{(2)}=dI/d\Omega\approx4/\Delta E_\gamma$ as a function 
of the angular frequency (see the second panel from the bottom, Fig.\ref{moi}). 
Here, $\hbar\Omega=E_{\gamma}/2$, $E_{\gamma}$ 
is the $\gamma$-transition energy between two neighboring states that differ on 
two units of the angular momentum $I$ and $\Delta E_{\gamma}$ is the difference 
between two consecutive $\gamma$-transitions.
At the transition point $\Im^{(2)}$ wildly fluctuates with a 
huge amplitude in $^{156}Dy$,  whereas these fluctuations are quite mild in 
$^{162}Yb$. This behaviour can be understood by virtue of the expansion of the 
microscopic moment of inertia at small rotational frequency \cite{bel}
\begin{equation}
\label{ib}
{\Im}\approx {\cal J}_{IB}+{\cal J}_M + 
\sum_i \frac{\partial I}{\partial \beta_i}\frac{\partial {\beta_i}}{\partial \Omega}
+...
\end{equation}
The Inglis-Belyaev moment of inertia, ${\cal J}_{IB}$, neglects a residual two-body
interaction between quasiparticles. 
Its behaviour is similar to the one of the kinematical moment of inertia, 
$\Im^{(1)}_{1}$. 
The second term ${\cal J}_M$ is a Migdal moment
of inertia \cite{mig}, resulting from the effect of rotation on the residual two-body
interaction and, in particular, on the pairing interaction. 
In our calculations the pairing gaps change smoothly in accord with the 
phenomenological prescription, Eq.(\ref{17}). The third term describes 
the variation of the self-consistent mean field, namely, the change of the deformation 
$(\beta_{1,2}\equiv \beta, \gamma$) under rotation. 
Therefore, a drastic change of the mean field configuration 
(namely, $\gamma$ -deformation) in $^{156}$Dy 
(see Fig.\ref{phatr}, top panels) explains large fluctuations of the dynamical 
moment of inertia at the transition point. 
In contrast, the smooth behaviour of the function $F$ at the transition point 
(see Fig.\ref{phatr}, bottom panels) implies a 
small amplitude of fluctuations of the dynamical moment of inertia in $^{162}$Yb.
The magnitude of fluctuations can be traced by means of the ratio 
$\Pi={\Im}^{(2)}_1/{\Im}^{(1)}_1$ as a function of the rotational frequency 
(see the bottom panel in Fig.\ref{moi}).  While this ratio is about
$\Pi\sim 2$ at the transition point in $^{162}$Yb, in $^{156}$Dy it is much
larger $\Pi\sim 8$. Referring to above analysis of the mean field solutions,
we can formulate the empirical rule to detect the order of the quantum 
shape-phase transition at the backbending: {\it if the ratio 
$\Pi={\Im}^{(2)}/{\Im}^{(1)}>>1$ at the transition point, the shape transition
can be associated with a first order phase transition}.

The nature of the backbending becomes more evident by dint
of the RPA analysis presented below. 
For a completeness, we also compare the Thouless-Valatin moment of inertia
(the definition is presented in Sec.III) with the dynamical moment of inertia 
$\Im^{(2)}_1$ calculated in the mean field approximation.
Notice, that all three terms in Eq.(\ref{ib}) are 
included into the Thouless-Valatin moment of inertia, ${\Im}_{RPA}={\Im}$, 
calculated in the RPA (see, for example, the discussion in Ref.\onlinecite{n2}
for the exact model without pairing), 
which is valid at large rotational frequencies as well.
One can observe a remarkable agreement between
experimental, mean field and CRPA results.
The agreement between  the mean field and CRPA results
confirms a good accuracy of our mean field calculations.
This agreement, on the other hand, demonstrates a 
validity of our CRPA approach in the backbending region, at least,
for the description of the spectrum.

\section{Quadrupole Collective excitations}

\subsection{Quasiparticle RPA in rotating systems}

In this section we discuss the RPA results to get a further insight into the 
backbending mechanism.
By means of a generalized Bogoliubov transformation, we express the Hamiltonian
given by Eq. (\ref{3}) in terms of quasiparticle
creation ($\alpha^\dagger_i$) and annihilation ($\alpha_i$) operators.
We then face with the RPA equations of motion, written  in the form \cite{KN,JK}
\begin{eqnarray}
\label{10}
[H_{\Omega}, P_{\nu}] \,&=&\, i\,\hbar \omega_{\nu}^2\,X_{\nu}, \qquad
[H_{\Omega}, X_{\nu}] \,=\, -i\,\hbar \,P_{\nu}, \nonumber\\
&&[X_{\nu}, P_{\nu'}]\,=\,i \hbar \delta_{\nu\,\nu'},
\end{eqnarray}
where $X_{\nu}$, $P_{\nu}$ are, respectively, the collective coordinates
and their conjugate momenta.
The solution of the above equations yields the RPA
eigenvalues $\hbar \omega_{\nu}$
and eigenfunctions
\begin{eqnarray}
|\nu> \,= \,O_{\nu}^\dagger |RPA> &=&
\frac{1}{\sqrt{2}}\,\Bigl(\,\sqrt{\frac{\omega_{\nu}}{\hbar}}
\,X_{\nu} \,-\,
\frac{i}{\sqrt{\hbar \omega_{\nu}}}\,\hat{P}_{\nu} \,\Bigr)
|RPA>\nonumber\\
&=& \sum_{\mu} \Bigl(\psi_{\mu}^\nu b^\dagger_{\mu} - \phi^\nu_{\mu}
b_{\mu} \Bigr) |RPA>,
\label{amp}
\end{eqnarray}
where $\mu=({k{\bar l}})$ or $({kl},{\bar k}{\bar l})$.  Here, 
the boson-like operators
$b^{+}_{k\bar{l}}=\alpha^{+}_{k}\alpha^{+}_{\bar{l}},
b^{+}_{kl}=\alpha^{+}_{k}\alpha^{+}_{l},
b^{+}_{\bar{k}\bar{l}}=\alpha^{+}_{\bar{k}}\alpha^{+}_{\bar{l}}$
are used.
The first equality introduces the positive signature boson,
while the other two determine the negative signature ones.
These two-quasiparticle operators are treated in the
quasi-boson approximation (QBA) as an elementary bosons, i.e.,
all commutators between them are approximated by their
expectation values with the uncorrelated HB vacuum \cite{RS}.
The corresponding commutation relations can be found in Ref.\onlinecite{KN}.
In this approximation any single-particle operator
$\hat{F}$ can be expressed as
$\hat{F} = \langle F \rangle + \hat{F}^{(1)}+\hat{F}^{(2)}$
where the second and third terms are linear and bilinear order terms in
the boson expansion. We recall that in the QBA one includes
all second order terms into the boson Hamiltonian such
that $(\hat{F}-\langle F \rangle)^2=\hat{F}^{(1)}\hat{F}^{(1)}$.
The positive and negative signature boson spaces are not mixed,
since the corresponding operators commute and
$H_{\Omega}=H_\Omega(r=+1) + H_\Omega(r=-1)$.
Consequently, we can solve the eigenvalue equations (\ref{10})
for $H_{\Omega} (+)$ and $H_{\Omega} (-)$, separately.

The symmetry properties of the cranking Hamiltonian yield
\begin{equation}
\label{spurious1}
[\, H_{\Omega}(+) \,,\,N_{\tau=n,p}\,]_{RPA}\, =  \,0,
\,\,\,\,[H_{\Omega}(+)\,,\,I_1\,]_{RPA}=0.
\end{equation}
The presence of the cranking term in Hamiltonian (\ref{3})
leads to the following equation
\begin{equation}
[H_\Omega(-),\Gamma^{\dagger}]=\hbar \Omega \Gamma^{\dagger},
\label{rot}
\end{equation}
where  $\Gamma^{\dagger}=(I_2 + i I_3)/\sqrt{2 \langle I_1 \rangle}$
and $\Gamma = (\Gamma^{\dagger})^{\dagger}=
(I_2 - i I_3)/\sqrt{2 \langle I_1 \rangle}$
fulfill the  commutation relation
\begin{equation}
[\Gamma, \Gamma^{\dagger}]=\hbar.
\end{equation}
According to Eqs.(\ref{spurious1}),
we have two Nambu-Goldstone modes, one is
associated with the violation of the conservation law for a 
particle number, the other is a positive signature
zero frequency rotational solution associated with
the breaking of spherical symmetry.
Eq.(\ref{rot}), on the other hand, yields
a negative signature  redundant solution
of energy $\omega_{\nu}=\Omega$, which describes
a collective rotational mode arising from the symmetries broken
by the external rotational field (the cranking term).
Eqs. (\ref{spurious1}) and (\ref{rot}) ensure
the separation of the spurious or redundant solutions
from the intrinsic ones. They would be automatically
satisfied if the single-particle basis was generated by means of a
self-consistent HB calculation.
As we shall show, they are fulfilled with a good accuracy also
in our, not fully self-consistent, HB treatment.

We recall that the yrast states possess the positive signature
quantum number. Obviously,  SSB effects of the mean field are related
to Nambu-Goldstone modes of the same signature.
Therefore, in this paper our analysis is focused upon positive
signature RPA excitations.

The positive signature Hamiltonian consists of the following terms
\begin{eqnarray}\nonumber
&&\hat H_{\Omega}[r =+1]=\sum_{ij}E_{i\bar{j}}b^{+}_{i\bar{j}}b_{i\bar{j}}-
\sum_{\tau = N,P}G_{\tau}\hat{P}_{\tau}^{(1)^{+}}\hat{P}_{\tau}^{(1)} \nonumber\\
&-&1/2\sum_{T =0,1}\,\kappa_{0}[T]
(\tilde M ^{(1)}[^{T}_{r=+1}])^2
-\frac{1}{2}\kappa_{\sigma}\sum_{T=0,1}
(\tilde s_1 ^{(1)}[^{T}_{r=+1}])^2
\nonumber\\
&-&\frac{1}{2}\sum_{T=0,1}\,\kappa_{2}[T]
\sum_{\mu=0,1,2} (\tilde Q^{(1)}_\mu [^{T}_{r=+1}])^2
\label{hps}
\end{eqnarray}
Here,  $E_{i\bar{j}} = \varepsilon_{i} + \varepsilon_{\bar j}$ are
two-quasiparticle energies and the definition of matrix elements of
the operators involved in the Hamiltonian (\ref{hps}) can be found in 
Ref.\onlinecite{JK}. The solution of the equations of motion, Eq.(\ref{10}), leads
to the following determinant of the secular equations
\begin{equation}
\label{det}
{\cal F}(\omega_{\nu}) =  \det~({\bf R}- \frac{\bf 1}{2 c}~),
\end{equation}
with a dimension $n=12$ and $c=\kappa_0, \kappa_2$ or $G_{\tau}$.
The matrix elements
$R_{km}(\omega_{\nu}) =
\sum_\mu q_{k,\mu }q_{m,\mu } C_\mu ^{km}/(E_\mu^2 -\omega_{\nu}^2)$
involve the coefficients $ C_{\mu}^{km} = E_{\mu }$ or $\omega_{\nu}$  
for different combinations of matrix elements $q_{k,\mu}$
(see details in Refs.\onlinecite{KN,JK}). The zeros of the function
${\cal F}(\omega_\nu)=0$
yield the CRPA eigenfrequencies $\omega_\nu$.
Once the RPA solutions are found, the Hamiltonian (\ref{hps}) can be written
in terms of the collective modes ($\hat{\mathit{X}_{\nu}}$,
$\hat{\mathit{P}_{\nu}}$)
\begin{eqnarray}
\nonumber
\hat{H}_{\Omega}[r] &=& \frac{1}{2} \sum_{\nu}\left( \hat{\mathit{P}}^{2}_{\nu}+
\omega^{2}_{\nu}\hat{\mathit{X}}^{2}_{\nu}\right) =
\sum_{\nu(\omega_{\nu}\neq 0)} \hbar \omega_{\nu}
\left( Q^{+}_{\nu}Q_{\nu} + \frac{1}{2} \right)  \nonumber\\
&+& \frac{1}{2} g_{I_{1}} \hat{\mathit{I}}^{(1)^{2}}_{1} + \frac{1}{2}
\sum_{\tau} g_{N_{\tau}} \hat{\mathit{N}}^{(1)^{2}}_{\tau}
\end{eqnarray}
where $g_{I_{1}}=1/\Im_{RPA}$, $g_{N_{\tau}}$ are the Thouless-Valatin moment of 
inertia and the nucleus mass, respectively, determined in the uniform rotating 
(UR) frame. The general derivation of the mass parameters $g_{I_{1}}$ and $g_{N_{\tau}}$
can be found in Refs.\onlinecite{KN,29}. In particular, for the moment of inertia
we have
\begin{equation}
\Im_{RPA}= \frac{det A}{det B}
\label{TV}
\end{equation}
where $A_{kl}$ is the matrix (the dimension  $n=10$)
given by the part of the matrix ${\cal F}_{kl} (\omega=0) $
corresponding to the s.p. operators $k$, and $l$ involved in
$\hat H[r=+1]$ with the structure  $(b^{+}_{i\bar{j}} +b_{i\bar{j}})$ type.
The matrix $B_{kl}$ (with
the dimension  $n=11$) is given by matrix $A_{kl}$ with additional one column
and row involving $0$ and sums $\pm \sum_{\mu = i\bar{j}}
\frac{f^{(k)}_{\mu}J^{(1)}_{\mu}}{E_{\mu}}$, $(k = 1,...,8)$ where
$J^{(1)}_{i\bar{j}}$ are quasiparticle matrix elements of the operator
$\hat I_{1}$ (see Ref.\onlinecite{29} for details).

Transition probabilities for $X\lambda$ transition $|I\nu> \rightarrow |I'\nu'>$
between two high-spin states is given by expression
\begin{eqnarray}
&&B(X\lambda; I\nu \rightarrow I'\nu') = \\
&&(I\,I\,\lambda\,\mu_{1}\,|\,I'\,I')^{2} |
<\nu'|\hat{\cal M}^{(1)}(X\lambda; \mu_{1}=I' - I)|\nu>|^{2}\nonumber
\end{eqnarray}
At high spin limit ($I >> \lambda$, $I' >> \lambda$), this expression takes 
the form  \cite{mar1} 

\begin{eqnarray}
\label{be2}
&&B(X\lambda; I\nu \rightarrow I' yr) \simeq \\
&&|<RPA|\left[ Q_{\nu},\hat{\cal M}^{(1)}
(X\lambda; \mu_{1}=I' - I)\right]|RPA>|^{2}\nonumber
\end{eqnarray}
for the transition from one phonon states into the yrast line states. Here,
$\hat{\cal M}^{(1)}(X\lambda\mu_{1})$ is the linear boson part of the
corresponding transition operator of type $X$, multipolarity $\lambda$ and the
projection $\mu_{1}$ onto the rotation axis (the first axis of the UR system).
The commutator in (\ref{be2}) can be easily expressed in terms of phonon amplitudes
$\psi^{(\nu)}_{\mu}$, $\phi^{(\nu)}_{\mu}$ ($\mu = i\bar{j}$ or $ij, \bar{i}\bar{j})$
and $\mid RPA \rangle $ denotes the RPA vacuum (yrast state) at
the rotational frequency $\Omega$.
The multipole operator in the rotating frame
is obtained from the corresponding one in the laboratory frame
according to the prescription \cite{mar1}
\begin{equation}
{\cal M}(X\lambda \mu_{1}) =
\sum_{\mu_{3}} \,\,{\cal D}^\lambda_{\mu_{1} \mu_{3}}
(0,\frac{\pi}{2},
0) \,\,{\cal M} (X \lambda \mu_{3}).
\label{bdf}
\end{equation}

Taking into account  that $<\nu|\hat M^{(E)}_{2\mu_{3}=0,2}[+]|\nu> =
\langle|\hat M^{(E)}_{2\mu_{3}=0,2}[+]|\rangle$ holds in the first RPA order  and
Eq.(\ref{bdf}), we have
\begin{eqnarray}
\label{tayr}
&&B(E2; I \nu \rightarrow I - 2 \, \nu) = \left| <\nu|\hat{\cal M}^{(1)}(E2;
\mu_{1}=2)|\nu> \right|^{2} \nonumber\\
&=&\frac{1}{2} \left| \frac{\sqrt{3}}{2}
\langle|\hat{M}^{(E)}_{2 \mu_{3}=0}[+]|\rangle - \frac{1}{2}
\langle|\hat{M}^{(E)}_{2 \mu_{3}=2}[+]|\rangle \right|^{2}
\end{eqnarray}

Using the $\beta$~-~$\gamma$ deformation parameterization, Eq.(\ref{Hartree}), and
the oscillator value for the isoscalar quadrupole constant,
$\kappa_{2}[0] \approx \frac{4\pi}{5}\frac{m\omega^{2}_{0}}{<r^{2}>}
\approx \frac{4\pi}{3A} m\omega^{2}_{0} R^{2}$ ($R=1.2A^{1/3} fm$), we obtain
an approximate expression
\begin{eqnarray}
\label{tbyr}
B(E2; I\, yr\, \rightarrow \,&& I -2\, yr) \approx \nonumber\\ 
&&\frac{1}{2}
\frac{9e^{2}Z^{2}}{16\pi^{2}} R^{4}\beta^{2} \cos ^{2}(\frac{\pi}{6} - \gamma )
\end{eqnarray}
Here, we use $\hat{M}^{(E)}=(eZ/A) \hat{M}$. From this expression one can 
conclude that the change of the $\gamma$-deformation from
zero to $\pi/6$ will increase the transition probability. The negative sign of 
$\gamma$-deformation 
results in the decrease of this probability. As we will see below, 
this expression is useful for the analysis of the experimental data.

\subsection{Determination of the residual interaction strengths}

\begin{figure}[ht]
%\centerline{\psfig{figure=fig10.eps,width=3.2in,clip=}}
\includegraphics[height = 0.3\textheight]{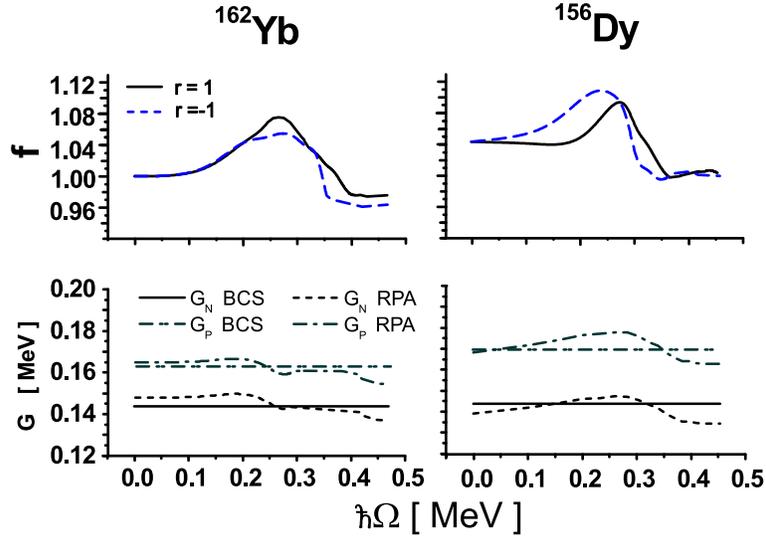}
\caption{
(Color online) The ratio $f=\kappa_{\it cal}/\kappa_{\it osc}$ between actual 
and oscillator isoscalar quadrupole strength constants  for the positive and 
negative signature is displayed on the upper panel.
On the lower panel, the  actual  and the BCS pairing constants $G_{N}$ and $G_{P}$ 
(see the definition in the text) are shown.
All actual values used in the calculations are obtained
as to fulfill  the RPA equations, Eqs.(\ref{spurious1})-(\ref{rot}),
for the spurious or redundant modes.
The results correspond to the calculation \textbf{I}, when
the term $\hat h_{add}$ (\ref{4a}), that restores the local
Galilean invariance broken in the Nilsson potential, is included.
}
\label{const}
\end{figure}

In order to determine the strength $\kappa$ of the monopole and
quadrupole interactions,  we start with the oscillator
values \cite{ds}
\begin{equation}
\kappa_{\lambda}[0]=
\frac{4\pi}{2 \lambda +1} \frac{m\omega_0^2}{A<{\tilde r}^{2\lambda -2}>},
\,\,\,\,\,\,\,\kappa_{\lambda}[1]=-\frac{\pi V_1}{A<{\tilde r}^{2\lambda}>}\,.
\label{HOkappa}
\end{equation}
For instance, the isoscalar  strength follows
from enforcing the Hartree self-consistent conditions.
We  then change slightly these   strengths and
the pairing interaction constants $G_{\tau}$
at each rotational frequency, while keeping constant the
$\kappa[1]/\kappa[0]$ ratio ($\kappa_{0}[1]/ \kappa_{0}[0] \approx -18$,
$\kappa_{2}[1] / \kappa_{2}[0] \approx -3.6$), so as to fulfill
the RPA equations (\ref{spurious1})-(\ref{rot}) for the spurious or redundant
modes.

 The rotational dependence of these parameters is relatively weak.
 The ratio between actual and oscillator value for the isoscalar quadrupole
 constant is displayed in Fig.\ref{const}.
 For comparison, the
BCS values $G_{\tau}$, $G_{N} \approx 22.5/A$ MeV,
$G_{P}\approx 26.5/A$ MeV, obtained from the systematic of pairing gaps
(see Ref.\onlinecite{33}), are shown in Fig.\ref{const} as well. Surprisingly, 
the difference between actual values and the latter ones is mild. However, 
the determination of $\kappa_{2}[0]$ and $G_{\tau}$ is a tedious task, since a 
tiny change of the strength parameters leads to large  shifts in energies of 
spurious modes. The constants so determined differ from the
HO  ones by 5-10\% at most. For the spin-spin interaction,  we use the
generally accepted strengths \cite{34}
$$
\kappa_{\sigma}[0]\,=\,\kappa_{\sigma}[1]\,=\,- 28 \,\frac{4\pi}{A}
\, MeV
$$
for all rotational frequencies. The spin-spin interaction does not influence the 
position of Nambu-Goldstone modes and, therefore, does not play any  role in the 
self-consistent determination of the quadrupole strength constants.
Finally, we adopted bare charges to compute the $E2$
strengths and a quenching factor $g_s = 0.7$ for the
spin gyromagnetic ratios to compute the $M1$ strengths.

By using the above set of parameters, it was possible not only
to separate the spurious and rotational solutions
from the intrinsic modes,
but also to reproduce the experimental dependence
of the lowest $\beta$- and $\gamma$-bands on the rotational frequency
$\hbar \Omega$ and,
in particular, to observe the
crossing of the $\gamma$- band with the ground band
in correspondence with the onset
of triaxiality.

Concluding the discussion about our calculation scheme, we should mention
some drawbacks of our approach. One of the disadvantage is 
the CRPA breaks down at the transition point when $\Delta_p$ or $\Delta_n$ vanish 
\cite{mar1}. We have avoided this problem by means of the phenomenological 
prescription for the rotational dependence of the pairing gap (see Section II). 
In principle, projection methods should be used in the transition region in order 
to calculate transition matrix elements. In the case of the phase transition of the 
first order we run into the problem of the formidable overlap integral.
A theory of large amplitude motion would provide a superior means to solve this 
problem. 

However, in contrast to the standard RPA calculations,   
where the residual strength constants are fixed for all values of $\Omega$ 
(see e.g. \cite{mat1,mat2}), we determine the strength constants
for each value of $\Omega$ by the requirement of the validity of the 
conservation laws. This enables us to overcome the instability 
of the RPA calculations at the transition region, at least, for
the excitations (see also discussion for quantum dots in Ref.\onlinecite{llor1}).
Although the amplitudes $\phi^{(\nu)}_{\mu}$ (see Eq.(\ref{amp})) 
are higher for the RPA modes in the transition
region than in other regions, the relation  
$|\phi^{(\nu)}_{\mu}|<|\psi^{(\nu)}_{\mu}|$ is still valid
to hold the QBA. And least but not last,
the CRPA becomes quite effective at high spins, when the pairing 
correlations are suppressed.

\subsection{Positive signature excitations in the rotating frame}

In this section we analyse the spectral and decay properties of the
positive signature states. Transition probabilities represent the most
challenging task, since they are sensitive about the wave function
structure of the considered state and, therefore, exhibit hidden drawbacks of
a theory.

\begin{figure}[ht]
%\centerline{\psfig{figure=fig11.eps,width=2.6in,clip=}}
\includegraphics[height = 0.28\textheight]{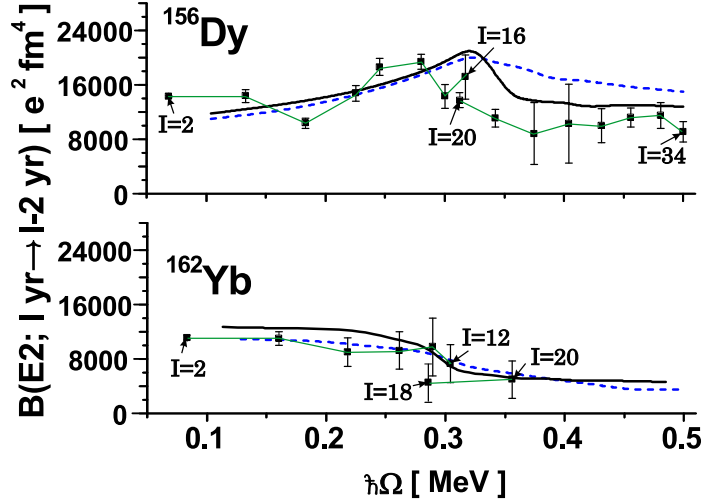}
\caption{
(Color online) Reduced transition probabilities 
$B(E2; I \, yr \, \rightarrow \, I-2 \, yr)$
along the yrast line. Experimental data (filled squares) 
are connected with a thin line to  guide
eyes. The results of calculations by means of Eq.(\ref{tayr}) and of 
Eq.(\ref{tbyr}) are connected by solid and dashed lines, respectively.
}
\label{beyr}
\end{figure}

Experimental values of $B(E2, \, I \, \nu \, \rightarrow \, I^{\prime} \, yr)$
are deduced from the half life  of the yrast states \cite{nndc} using
the standard, long wave limit expressions
$B(E2, i \, \rightarrow \, f) =\
P(i\, \rightarrow \, f)/(1.223\times10^{9}E^{5}_{\gamma}) 
\,\, (e^{2}$ fm$^{4})$ \cite{BM1,RS}. Here, the transition energy 
$E_{\gamma}$ is in MeV and the absolute transition 
probability $P(i\, \rightarrow \, f)=\ln 2 /T(i\, \rightarrow \, f)$ is the
related to the half life $T(i\, \rightarrow \, f)$ (in seconds).
For $^{156}$Dy the
yrast energies and corresponding half lifes are known up to the momentum
$I\approx 40 \hbar$. For $^{162}$Yb yrast energies are observed up to
$I \approx 28 \hbar$ but half lifes were measured only up to $I \approx 20\hbar$.
While the cranking approach should be complemented with a projection
technique in the backbending region due large fluctuation of the angular
momentum (cf Ref.\onlinecite{RS}), its validity becomes much better at high spins.
The agreement between calculated  and experimental values
of intraband $B(E2)$ transitions along the yrast line
is especially good after the transition point.

\begin{figure}[ht]
%\centerline{\psfig{figure=fig12.eps,width=3.6in,clip=}}
\includegraphics[height = 0.35\textheight]{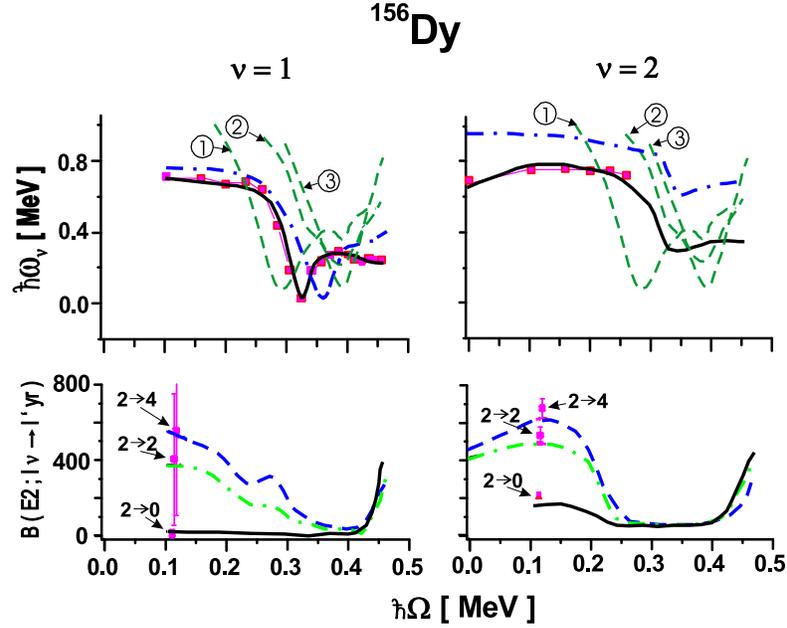}
\caption{
(Color online) $^{156}$Dy. 
Experimental (filed square) and calculated
excitation energies in the rotating frame
$\hbar \omega_{\nu}=R_{\nu}(\Omega)-R_{yr}(\Omega)$  (upper panel)
and reduced transition probabilities
$B(E2; I \, \nu \, \rightarrow \, I' \, yr)$ (lower panel) for two
lowest RPA solutions, $\nu=1$ (left) and $\nu=2$ (right),
as a function of the angular frequency $\hbar \Omega$.
The results for excitations, calculated with and without the term $\hat h_{add}$,
Eq. (\ref{4a}),
are connected by solid and dash-dotted lines, respectively.
The experimental values for the transitions are indicated by arrows.
The tansitions with $\Delta I=2,0,-2$ are connected by solid, dash-dotted and 
dashed lines, respectively.
Two-quasiparticle energies, indicated by arrows with numbers 1, 2, and 3 in circles,
originate from two s.p. Nilsson
states ($n \bar n \,\frac{3}{2}[651]\,\,\frac{3}{2}[651]$),
($p \bar p \, \frac{7}{2}[523]\,\,\frac{7}{2}[523]$),
and ($n \bar n \,\frac{1}{2}[530]\,\,\frac{5}{2}[523]$) at $\Omega=0$ and
$\Delta_{\tau}=0$.}
\label{dy}
\end{figure}

\begin{figure}[t]
%\centerline{\psfig{figure=fig13.eps,width=3.6in,clip=}}
\includegraphics[height = 0.35\textheight]{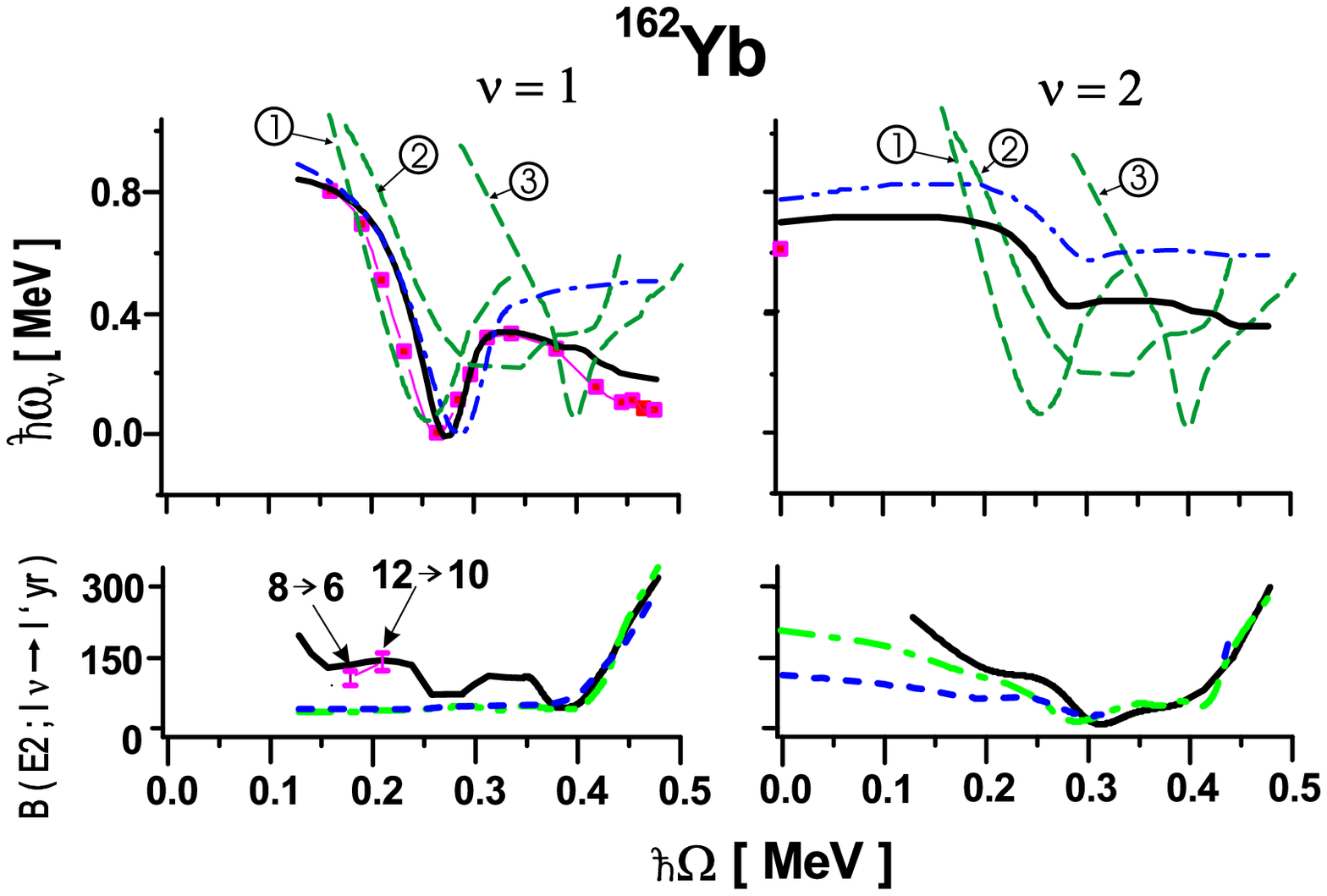}
\caption{
(Color online) $^{162}$Yb.
The same as for Fig.\ref{dy}.
Two-quasiparticle energies, indicated by arrows with numbers 1, 2, and 3 in circles,
originate from two s.p. Nilsson states ($n \bar n \,\frac{3}{2}[651]\,\,\frac{3}{2}[651]$),
($n \bar n \,\frac{3}{2}[521]\,\,\frac{3}{2}[521]$),
and ($p \bar p\, \frac{7}{2}[523]\,\,\frac{7}{2}[523]$)
at $\Omega=0$ and $\Delta_{\tau}=0$.}
\label{yb}
\end{figure}

Experimental data are compared with
the results of calculations (a) by means of Eq.(\ref{tayr})  and
calculations (b) by means  of Eq.(\ref{tbyr}) (see Fig.\ref{beyr}).
In the calculations (a)  we use the mean field
values for the quadrupole operators. In the calculations
(b) these values have been replaced by the deformation parameters via
Eq.(\ref{Hartree}) and  we use the oscillator value for the quadrupole strength 
constant. The calculations (a) evidently manifest the backbending effect obtained 
for the moments of inertia (see Fig.\ref{moi})
at $\hbar \Omega_c \approx 0.25,\, 0.3$ MeV for $^{162}$Yb and $^{156}$Dy, 
respectively. Thus, the use of the self-consistent
expectation values $\langle |\hat M^{(E)}_{2\mu_{3}}[+]|\rangle$ is crucial
to reproduce the experimental behaviour of the yrast band decay.
The calculations (b) (Eq.(\ref{tbyr})) reproduce the experimental data with
less accuracy. However, these results also catch on the correlation
between the sign  of the $\gamma$-deformation and the
behaviour of the transition probability.
The onset of the positive (negative) values of $\gamma$-deformation lead
to the increase (decrease) of the transition probability along the yrast line.
This fact nicely correlates with the experimental data.

To analyse experimental data
on low lying excited states near the yrast line
we construct the Routhian function for each rotational band $\nu$
($\nu = yrast, \beta, \gamma, ...$)
\begin{equation}
R_{\nu}= E_{\nu}(\Omega) - \hbar \Omega I(\Omega), \quad \quad \Omega(I)=
\frac{E_{\nu}(I+2)-E_{\nu}(I)}{2}
\end{equation}
and define the experimental excitation energy in the rotating frame
$\hbar \omega_{\nu}^{\it exp} = R_{\nu}(\Omega) - R_{yr}(\Omega)$ \cite{n87}. 
This energy can be directly compared with the corresponding 
solutions $\hbar \omega_{\nu}$ of the RPA secular equations for a given 
rotational frequency $\Omega$. We remind that our vacuum states are 
adiabatic quasiparticle configurations that correspond to the 
occupied orbitals at a given rotational frequency.

The first lowest positive signature RPA solutions
($\nu = 1$) create an excited band built on the yrast line. In both nuclei,
this band is close to the experimental $\gamma$-excitations with even spin at
small rotational frequency $\Omega$ (see Figs.\ref{dy}, \ref{yb}). In the 
low-spin region, the transitions $B(E2;\,I\nu=1 \,\rightarrow \, I' yr)$ are 
large and  RPA solutions exhibit a strong collective nature of the $\gamma$ - band
(Weisskopf units are $B(E2)_W \,=\,49.9,
\,52.5 \,e^2 fm^4$ for $^{156}Dy$, $^{162}Yb$, respectively) in both nuclei.
The collective character of the low-spin part of the $\gamma$ - bands
is manifested  in the phonon structure that is composed
by a few two-quasiparticle components (see below and Tables I,II).

In $^{156}$Dy the RPA results,
obtained with the term $\hat h_{add}$ (\ref{4a}), reproduce quite well the rotational
behaviour of the lowest excitations associated with the collective $\gamma$-vibrations. 
We also obtain a good agreement with the experimental value of the critical 
rotational frequency $\hbar \Omega_{cr}\approx 0.324$ MeV, at
which this mode disappear in the rotating frame. This results 
is very close to the critical rotational frequency 
$\hbar \Omega_c \approx 0.3$ MeV (see Fig.\ref{phatr}), where the backbending occurs 
(see Eq.(\ref{man}) and discussion in Sec. II). 
In accordance with the phenomenological theory of the first order phase
transitions discussed in Section II, this soft mode creates a shape transition
from the axial to the nonaxial shapes.
We emphasize that without the term $\hat h_{add}$ the transition point is
located considerably higher than the experimental one.
In $^{162}$Yb, the results with and without this term are very
close (see Fig.\ref{yb}). In contrast to $^{156}$Dy,  in $^{162}$Yb the crossing,
is determined by a single two-quasiparticle state (see Table I at $I=10\hbar$). 
The  term $\hat h_{add}$ results in a
collective effect to the mean field solution, while its contribution
to a single, two-quasiparticle energy $E_{\mu}$ is weak.
As discussed in Sec. II, the effect of this term is also reduced due to the 
alignment.

\begin{table*}
\caption{The structure of the $\nu=1$ positive signature phonon in $^{162}Yb$.}
\begin{tabular}{|c|c|c|c|}
\hline
& & & \\
$I=2\,$
& $n \bar n \,\frac{3}{2}[651] \,\,\frac{1}{2} [660] \,49$\%
& $n \bar n \,\frac{3}{2}[651] \,\,\frac{3}{2} [651] \,\,40$\%
& $p \bar p \,\frac{7}{2}[523] \,\,\frac{11}{2} [505] \,\,13$\%\\
& & & \\
\hline
& & & \\
$I=6\,$
& $n \bar n \,\frac{3}{2}[651] \,\,\frac{3}{2} [651] \,\,72$\%
& $n \bar n \,\frac{3}{2}[651] \,\,\frac{1}{2} [660] \,\,20$\%
& $p \bar p \,\frac{7}{2}[523] \,\,\frac{7}{2} [523] \,\,\,4$\%\\
& & & \\
\hline
& & & \\
$I=10$
& $n \bar n \,\frac{3}{2}[651] \,\,\frac{3}{2} [651] \,\,96$\%
& $p \bar p \,\frac{7}{2}[523] \,\,\frac{7}{2} [523] \,\,\,2$\%
& $n \bar n \,\frac{1}{2}[651] \,\,\frac{1}{2} [660] \,\,\,1$\%\\
& & & \\
\hline
& & & \\
$I=14$
& $n \bar n \,\frac{3}{2}[651] \,\,\frac{3}{2} [651] \,\,85$\%
& $n \bar n \,\frac{3}{2}[521] \,\,\frac{3}{2} [521] \,\,\,7$\%
& $p \bar p \,\frac{7}{2}[523] \,\,\frac{7}{2} [523] \,\,\,7$\% \\
& & & \\
\hline
& & & \\
$I=18$
& $n \bar n \,\frac{3}{2}[521] \,\,\frac{3}{2} [521] \,\,70$\%
& $p \bar p \,\frac{3}{2}[651] \,\,\frac{3}{2} [651] \,\,21$\%
& $p \bar p \,\frac{7}{2}[523] \,\,\frac{7}{2} [523] \,\,\,5$\%\\
& & &  \\
\hline
\end{tabular}
\label{tableyb}
\end{table*}

\begin{table*}
\caption{The structure of the $\nu=1$ positive signature phonon in $^{156}Dy$.}
\begin{tabular}{|c|c|c|c|}
\hline
& & & \\
$I=2\,$
& $n \bar n \,\frac{3}{2}[651] \,\,\frac{1}{2} [660] \,\,48$\%
& $p \bar p \,\frac{7}{2}[523] \,\,\frac{7}{2} [523] \,\,17$\%
& $n \bar n \,\frac{1}{2}[530] \,\,\frac{5}{2} [523] \,\,16$\%\\
& & & \\
\hline
& & & \\
$I=6\,$
& $n \bar n \,\frac{3}{2}[651] \,\,\frac{3}{2} [651] \,\,63$\%
& $p \bar p \,\frac{1}{2}[550] \,\,\frac{1}{2} [541] \,\,15$\%
& $n \bar n \,\frac{1}{2}[530] \,\,\frac{5}{2} [523] \,\,13$\%\\
& & & \\
\hline
& & & \\
$I=10$
& $n \bar n \,\frac{3}{2}[651] \,\,\frac{3}{2} [651] \,\,60$\%
& $p \bar p \,\frac{1}{2}[550] \,\,\frac{7}{2} [523] \,\,17$\%
& $n \bar n \,\frac{1}{2}[530] \,\,\frac{5}{2} [523] \,\,14$\%\\
& & &  \\
\hline
& & & \\
$I=14$
& $n \bar n \,\frac{3}{2}[651] \,\,\frac{3}{2} [651] \,\,65$\%
& $p \bar p \,\frac{7}{2}[523] \,\,\frac{7}{2} [523] \,\,18$\%
& $n \bar n \,\frac{3}{2}[530] \,\,\frac{5}{2} [523] \,\,13$\%\\
& & & \\
\hline
& & & \\
$I=18$
& $n \bar n \,\frac{3}{2}[651] \,\,\frac{3}{2} [651] \,\,71$\%
& $p \bar p \,\frac{7}{2}[523] \,\,\frac{1}{2} [541] \,\,12$\%
& $n \bar n \,\frac{1}{2}[530] \,\,\frac{5}{2} [523] \,\,13$\%\\
& & & \\
\hline
\end{tabular}
\label{tabledy}
\end{table*}

In Tables \ref{tableyb}, \ref{tabledy} we present the contribution
$n_{i\bar j}(\nu=1)=(\psi^{(\nu=1)}_{i\bar j})^2 -(\varphi^{(\nu=1)}_{i\bar j})^2$
of main two-quasiparticle components to the norm $\sum_{ij} n_{i\bar j}(\nu=1)\,=\,1$
of the RPA solution $\hbar \omega_{\nu=1}$ as a function of the angular momentum 
(rotational frequency). The collectivity is weaker in $^{162}$Yb
than in $^{156}$Dy. With the increase of the rotational frequency, in $^{162}$Yb
the phonon loses the collective nature and a single, two-quasiparticle neutron 
component ($i\bar j \,=\, n\bar n \, \frac{3}{2}[651] \, \frac{3}{2}[651]$) 
is dominant in the transition region. At the crossing point its weight reaches 
96\% and $B(E2;\,I\nu=1 \rightarrow I'yr)$ values falls down to their s.p. values
$B(E2)_W$.  The transition from axially symmetric to nonaxial shapes occurs due to
the alignment of this two-quasiparticle component along the axis of the collective 
rotation. There the SSB effects
display {\it a single-particle} mechanism.

In contrast with $^{162}$Yb,  in $^{156}Dy$ the phonon excitation 
$\hbar \omega_{\nu=1}$ remains collective
even in the backbending region (see Table \ref{tabledy} for $I\approx 14 \hbar$).
The $B(E2;\,I\nu=1 \rightarrow I'yr)$ values decrease with the increase of the
rotational frequency but  the phonon holds the collectivity  even  at
the transition point $\hbar \Omega_c=0.301$ MeV.
Although two-quasiparticle states align their angular momenta along the 
axis x (collective rotation), the axial symmetry persists till the transition point 
(see also discussion in Ref.\onlinecite{JK1}). The mode blocks a transition to
the triaxial shape.
At the transition point the phonon becomes anomalously soft 
and we obtain a change of the symmetry of the HB solution from axially symmetric 
to nonaxial shape. Thus, the SSB effect occurs due to {\it vanishing of the collective 
$\gamma$-vibrational excitations in the rotating frame}.

The second positive signature nonspurious RPA solution, $\hbar \omega_{\nu = 2}$
can be associated  with the $\beta$~-~band at small rotational frequencies.
For $^{156}$Dy the agreement between the experimental $\hbar \omega_{\beta}$
and the calculated $\hbar \omega_{\nu=2}$ is good (see Fig.\ref{dy}).
In $^{162}$Yb only one lowest level of $\beta$-band is experimentally
observed at $\hbar \Omega=0$ MeV and it is very close to
the calculated  value of the $\hbar \omega_{\nu=2}$. Our results
may be considered  as a theoretical prediction for the behaviour of 
the $\beta$-band at larger rotational frequencies.

\section{Summary}

We develop a practical method based on the cranked Nilsson potential
with separable residual interactions  for the analysis of the low-lying excitations
near the yrast line.  In contrast to previous studies of low-lying excitations
at high spins (cf Refs.\cite{mat1,mat2}), we pay a special attention
to the self-consistency between mean field results and description of
low-lying excitations in the RPA. We accounted for the $\Delta N=2$
coupling in generating the Nilsson states and included the Galilean
invariance restoring piece according to the prescription of Ref.\onlinecite{nak}. 
Moreover, we enforced the HB stability conditions, provided by
Eq.(\ref{new}), that yield deformation parameters very close to the
self-consistent values. Finally, we fixed the strength parameters of
the interaction so as to ensure the separation of the spurious modes from the
intrinsic excitations at each rotational frequency. This way we provide a reliable
approach to study the spontaneous breaking effects of continuous symmetries of
the rotating mean field. Note that even in self-consistent calculations with 
effective forces (Gogny or Skyrme type) this separation is not guaranteed due to
a finite size of the configuration space. One needs an extended configuration
space to ensure a good separation of a spurious contribution to the intrinsic 
wave function (see, for example, discussion in Ref.\onlinecite{llor1}).

We analyse the rotational properties of the yrast and low-lying positive signature
excitations in the transitional nuclei $^{156}$Dy and $^{162}$Yb undergoing the 
backbending. The agreement between our results  and experimental data is remarkable.
We obtain a simple expression, Eq.(\ref{tbyr}), for the reduced transition probability 
along the yrast line, that naturally explains the increase/decrease of $B(E2)$-transitions 
due to the shape transition from axially symmetric to nonaxial shapes with different sign 
of the $\gamma$-deformation of the rotating  mean field.
The magnitude of the B(E2) transition probability along the yrast line increases 
with the angular momentum and  drops down at the transition point in both nuclei.
We demonstrate a good agreement between the dynamical moment of inertia
calculated at the mean field level and the Thouless-Valatin moment of inertia 
calculated in the RPA in the realistic calculations. This is one of the stringent
test of the self-consistency between mean field and RPA calculations.
Our RPA analysis reveals a mechanism of the backbending
phenomena in rotating nuclei, which is caused by the disappearance of the
$\gamma$-vibrations in the rotating frame and following alignment of the
two-quasiparticle components of the $\gamma$-phonon.

We found that in axially symmetric nuclei in the ground state
two types of quantum phase transitions  may occur with rotation, which are 
associated with the backbending.
In $^{156}$Dy  we obtain that $\gamma$-vibrational excitations ($K=2$)
tend to zero in the rotating frame with the increase of the rotational frequency
$\Omega$, in close agreement with experimental data. Although two-quasiparticle 
states align their angular momenta along the axis x (collective rotation), 
the axial symmetry persists, since
the vibrational mode blocks a transition to the triaxial shape.
Near the transition point $\Omega_c$ there are two HB minima with different 
shapes: axially symmetric and strongly nonaxial.
At the transition point the phonon energy tends to zero and we obtain a change
of the symmetry of the HB solution from axially symmetric to nonaxial shape,
which corresponds to the backbending in $^{156}$Dy. 
A drastic change of the mean field configuration leads to large fluctuations of 
the dynamical moment of inertia at the transition point. 
The observed phenomenon resembles very 
much the structural phase transition discussed within the anharmonic 
Landau-type model in solid state physics \cite{krum}. We propose to consider  
{\it the onset of the $\gamma$-instability at the transition point 
associated with the backbending,
accompanied with large fluctuations of the dynamical moment of inertia, 
that exceed by few times the value of the kinematical moment of inertia, 
as a manifestation of a shape-phase transition of the first order}. 
In contrast with $^{156}$Dy, at the vicinity of the transition point $\Omega_c$ 
in $^{162}$Yb a single neutron two-quasiparticle component dominates ($\sim 96\%$) 
in the phonon structure. There the backbending is caused by  
the alignment of the neutron two-quasiparticle configuration. 
The energy $E_{\Omega}(\beta,\gamma)$  and the order parameter $\gamma$ 
(see Fig.\ref{phatr}) are smooth  functions in the vicinity of the transition point 
$\Omega_c$. The smooth behaviour of the 
energy at the transition point implies a  small amplitude of fluctuations of the 
dynamical moment of inertia. Extending the Landau-type theory for rotating nuclei, 
we proved that this transition can be associated with the second order quantum 
phase transition. 
Thus, {\it if the amplitude of the 
fluctuations of the dynamical moment of inertia at the transition point is of 
the same order of magnitude as the value of the kinematical moment of inertia, 
a backbending  may be associated with a 
quantum shape-phase transition of the second order}. 
We expect that different mechanisms of the backbending should affect 
wobbling excitations that attract a considerable attention nowadays. 
This is a subject of the forthcoming paper.

\section*{Acknowledgments}
We thank S. Frauendorf for useful discussions at early stage of this
analysis. We are grateful to R. Bengtsson for the discussion on properties 
of the quasiparticle spectra, F. D\"{o}nau and D. Sanchez for the fruitful remarks. 
This work is a part of the research plan MSM
0021620834 supported by the Ministry of Education of the Czech Republic.
It is also partly supported by Grant No.\ BFM2002-03241
from DGI (Spain). R. G. N. gratefully acknowledges support from the
Ram\'on y Cajal programme (Spain).


\begin{thebibliography}{99}
\bibitem{fot}
             according to the previous version of database \cite{nndc} 
             the energy level with $\hbar\omega=0.675$ MeV was assigned to
	     the state $K=2$. In this case 
	     $\hbar\Omega_{cr}=\hbar\omega_{K=2}/2=0.340$ MeV, which is closer to the 
	     critical rotational frequency $\hbar\Omega_{cr}=0.301$ MeV. 
	     In the updated version this energy level is considered 
	     as $K=0$ state. The resolution of this point could be useful to 
	     establish the relation between the critical rotational frequency and 
	     the vibrational frequency.
\end{thebibliography}
\end{document}